\documentclass[trackchanges,twocolumn]{aastex701}
\usepackage{textcomp}

\usepackage{graphicx}
\usepackage{amsmath}
\usepackage[varg]{txfonts}
\usepackage{xcolor}
\usepackage{xcolor}
\usepackage{tabularx}
\usepackage{array}
\usepackage{rotating}
\usepackage{makecell}
\usepackage{tablefootnote}
\usepackage{booktabs}
\usepackage{threeparttable}
\usepackage{siunitx}
\usepackage{soul}
\usepackage[utf8]{inputenc}
\usepackage{hyperref}
\usepackage{multirow}
\usepackage{chngcntr}
\counterwithout{figure}{section}
\def\oiiiab{[O\,{\sc iii}]$\lambda\lambda$4959,5007}
\def\oiiia{[O\,{\sc iii}]$\lambda$4959}
\def\oiiib{[O\,{\sc iii}]$\lambda$5007}
\def\halpha{H$\alpha$}
\def\hbeta{H$\beta$}
\def\niiab{[N\,{\sc ii}]$\lambda\lambda$6548,6584}
\def\siiab{[S\,{\sc ii}]$\lambda\lambda$6716,6731}
\def\niia{[N\,{\sc ii}]$\lambda$6548}
\def\niib{[N\,{\sc ii}]$\lambda$6584}

\def\totxrgs{187\,}
\def\oiiidp{55\,}

\def\niibptrobust{16\,}

\def\niiuplim{21\,}

\def\siibptrobust{13\,}

\def\dpbhmassmean{9.04\,}
\def\dpbhmasslow{7.71\,}
\def\dpbhmassup{11.05\,}
\def\dpbhmassmeanerr{0.11\,}
\def\spbhmassmean{9.08\,}
\def\spbhmasslow{7.67\,}
\def\spbhmassup{11.18\,}
\def\spbhmassmeanerr{0.12\,}
\def\dpradiolummean{26.08\,}
\def\dpradiolumlow{24.79\,}
\def\dpradiolumup{27.25\,}
\def\dpradiolummeanerr{0.01\,}
\def\spradiolummean{25.91\,}
\def\spradiolumlow{21.04\,}
\def\spradiolumup{27.62\,}
\def\spradiolummeanerr{0.02\,}

\usepackage{placeins}

\DeclareUnicodeCharacter{02BC}{'}

\begin{document}

\title{Origin of the double-peaked narrow emission-lines in the optical spectra of X-shaped Radio Galaxies}

\author[orcid=0009-0005-5992-6646,gname=Ghosh, sname='Prajnadipt']{Ghosh Prajnadipt}
\affiliation{Indian Institute of Astrophysics (IIA), Koramangala, Bangalore, \it{560034}; India}
\affiliation{National Centre for Nuclear Research, Pasteura 7, 02-093 Warsaw, Poland}
\email{prajnadiptghosh@gmail.com}

\author[orcid=0000-0002-5535-4186,gname=Ravi, sname='Joshi']{Ravi Joshi} 
\affiliation{Indian Institute of Astrophysics (IIA), Koramangala, Bangalore, \it{560034}; India}
\email{rvjoshirv@gmail.com}

\author[orcid=0000-0002-4439-5580,gname=Xiaolong,sname=Yang]{Xiaolong Yang}
\affiliation{Key Laboratory for Research in Galaxies and Cosmology, Shanghai Astronomical Observatory, Chinese Academy of Sciences, 80 Nandan Road, Shanghai 200030, People’s Republic of China}
\email{}

\author[orcid=0000-0001-8256-8887,gname=Yingkang,sname=Zhang]{Yingkang Zhang}
\affiliation{Key Laboratory for Research in Galaxies and Cosmology, Shanghai Astronomical Observatory, Chinese Academy of Sciences, 80 Nandan Road, Shanghai 200030, People’s Republic of China}
\email{}

\author[gname=Gopal-Krishna]{Gopal-Krishna}
\affiliation{UM-DAE Centre for Excellence in Basic Sciences (CEBS), Vidyanagari, Mumbai, 400098, India}
\email{}

\author[orcid=0000-0002-1029-3746,gname='Paul J.',sname=Wiita]{Paul J. Wiita}
\affiliation{Department of Physics, The College of New Jersey, 2000 Pennington Road, Ewing, NJ 08628-0718, USA}
\email{}

\author[orcid=0000-0002-7018-4349,gname=Dusmanta, sname='Patra']{Dusmanta Patra}
\affiliation{Department of Basic Science and Humanities, Gargi Memorial Institute of Technology, Baruipur, Kolkata – 700144, India}
\email{}

\author[orcid= 0009-0003-2002-7849,gname=Ankit,sname=Patel]{Ankit Patel}
\affiliation{Indian Institute of Astrophysics (IIA), Koramangala, Bangalore, \it{560034}; India}
\email{}

\author[orcid=0000-0002-2224-6664,gname=Arti,sname=Goyal]{Arti Goyal}
\affiliation{Obserwatorium Astronomiczne Uniwersytetu Jagiello\'nskiego, ul. Orla 171, 30-244 Krak\'ow, Poland}
\email{}

\author[orcid=0000-0001-9133-1005,gname=Gourab,sname=Giri]{Gourab Giri}
\affiliation{Istituto Nazionale di Astrofisica (INAF) – Istituto di Radioastronomia (IRA), via Gobetti 101, 40129 Bologna, Italy}
\email{}

\author[orcid= 0000-0003-0793-6066,gname=Santanu,sname=Mondal]{Santanu Mondal}
\affiliation{Indian Institute of Astrophysics (IIA), Koramangala, Bangalore, \it{560034}; India}
\email{}

\author[orcid=0000-0001-5824-1040,gname='Vibhore',sname=Negi]{Vibhore Negi}
\affiliation{Kavli Institute for Astronomy and Astrophysics, Peking University, Beijing 100871, Peopleʼs Republic of China}
\email{}

\author[orcid=0000-0002-4112-9607,gname='Marek',sname='We\.zgowiec']{Marek We\.zgowiec}
\affiliation{Obserwatorium Astronomiczne Uniwersytetu Jagiello\'nskiego, ul. Orla 171, 30-244 Krak\'ow, Poland}
\email{}

\author[orcid=0000-0003-3080-9778,gname=Katarzyna, sname='Ma{\l}ek']{Katarzyna Ma{\l}ek}
\affiliation{National Centre for Nuclear Research, Pasteura 7, 02-093 Warsaw, Poland}
\email{}

\author[orcid=0000-0002-7350-6913,gname='Xue-Bing',sname=Wu]{Xue-Bing Wu}
\affiliation{Kavli Institute for Astronomy and Astrophysics, Peking University, Beijing 100871, Peopleʼs Republic of China}
\affiliation{Department of Astronomy, School of Physics, Peking University, Beijing 100871, Peopleʼs Republic of China}
\email{}

\author[orcid=0000-0001-6947-5846,gname='Luis C.',sname=Ho]{Luis C. Ho}
\affiliation{Kavli Institute for Astronomy and Astrophysics, Peking University, Beijing 100871, Peopleʼs Republic of China}
\affiliation{Department of Astronomy, School of Physics, Peking University, Beijing 100871, Peopleʼs Republic of China}
\email{}

\begin{abstract}
    We investigate X-shaped radio galaxies (XRGs) with optical double-peaked narrow emission lines (DPNELs) as potential hosts of dual or binary supermassive black holes (SMBHs). Using available optical spectra from the SDSS and DESI surveys, we identify 55 DPNEL sources among a sample of 187 XRGs. We find that the occurrence of [O {\sc iii}] DPNELs in XRGs is $\sim$30\%, which is substantially higher than the $\sim$1\% observed in the general galaxy population. Using optical (involving \oiiib, \halpha, \hbeta, \niib, and \siiab) and mid-infrared diagnostic diagrams, we found that both components are predominantly AGN-powered, implying a $\sim$95\% probability of hosting dual AGN systems. In comparison, a stellar-mass-, color-, and redshift-matched control sample of non-XRG DPNEL galaxies shows a lower dual AGN fraction, which depends strongly on radio luminosity,  rising from ~25\% in radio-undetected to ~54\% in radio-detected galaxies, and reaching ~95\% in radio-bright XRGs and FR-II radio galaxies. Interestingly, more than 30\% of DPNEL XRGs have companion galaxies, supporting a merger-driven origin for these systems. Finally, we analyze parsec-scale radio structure in 7 XRGs using VLBA observations at 1.4, 4.3, and 7.6 GHz, resolving the core in only one source. Nevertheless, flat radio spectral indices, AGN-like emission-line properties, and radio-optical VLBA-Gaia positional offsets found in 5 cases together support the interpretation that XRGs are strong dual/binary AGN candidates.
\end{abstract}

\keywords{\uat{Galaxies}{573} --- \uat{Active Galaxies}{17} --- \uat{Active Galactic Nuclei}{16} --- \uat{Radio jets}{1347} --- \uat{Supermassive Black Holes}{1663} --- \uat{Radio Galaxies}{1343}}


\section{Introduction}
\label{sec:intro}
In the hierarchical model of galaxy evolution, mergers are thought to play a dominant role in the final evolutionary stages \citep{1978MNRAS.183..341W, 2000MNRAS.311..576K}. In the course of a merger, the central supermassive black holes (SMBHs) from each progenitor galaxy may form a close pair within the remnant galaxy \citep[e.g.][]{2006MmSAI..77..733K}. If both merging galaxies are gas-rich, substantial amounts of gas are funneled into their central regions, simultaneously triggering black hole accretion,  and initiating a central starburst \citep{1989Natur.340..687H, 2006ApJS..163....1H, 2015ApJ...806..219C}. This stage, characterized by two SMBHs actively accreting within the same galaxy at kiloparsec-scale separations, is known as the {\it dual} active galactic nuclei ({\it dual} AGN) phase.

As the merger progresses, dynamical friction causes the SMBHs to lose energy, bringing them from kiloparsec to parsec-scale separations \citep{1980Natur.287..307B}. Under favorable conditions, the two black holes enter a gravitationally bound orbital configuration \citep{2003ApJ...582..559V}, leading to the {\it binary} AGN phase. Eventually, gravitational wave emission dominates their energy loss, culminating in coalescence --- an event detectable by future space-based gravitational wave observatories such as the Laser Interferometer Space Antenna \citep[LISA; ][]{2017arXiv170200786A}.

Detecting dual and binary AGNs is the key to understanding the co-evolution of SMBHs, the fueling of AGN driven by galaxy mergers, and the formation of gravitational wave sources (see \citealt{2019NewAR..8601525D} and references therein). Cosmological and merger simulations suggest that the occurrence of dual AGNs correlates with the galaxy merger rate, which peaks around cosmic noon ($z \sim$ 1--3), where the corresponding dual AGN fraction is $\le 6\%$ \citep{2025MNRAS.536.3016P}. Further simulations indicate that for a sample of major and minor mergers hosting dual AGN, the fraction is 20-30\% and 1-10\% respectively \citep{2017MNRAS.469.4437C}.  Despite these theoretical expectations, dual AGNs remain quite rare from an observational perspective \citep{2017MNRAS.465.4772R,2021AJ....162..289Z}, conceivably due to challenges such as dust obscuration, AGN duty cycle, short lifetimes of the dual AGN phase, and limitations imposed by the required resolution of micro-arcseconds. In the recent Big Multi-AGN Catalog (Big MAC, \citealt{2025ApJS..281...25P}), a total of 156 confirmed dual AGN systems have been identified using techniques including optical spectroscopy, radio, and mid-infrared imaging; a further 4180 dual AGN candidates have been cataloged from the literature. In contrast, due to the requirement for parsec-scale resolution, only one confirmed binary AGN has been established (see below) \citep{2006ApJ...646...49R}, while approximately 1368 candidates are listed in Big MAC. Since direct detection methods such as high-resolution imaging in the X-ray \citep{2019ApJ...882...41H}, optical \citep{2014ApJ...780..106I, 2018ApJ...862...29L, 2023MNRAS.524.4482B}, radio bands \citep{2019MNRAS.484.4933R,2024ApJ...961..233S} and Integral Field Spectroscopy (IFS) \citep{2023A&A...679A..89P, 2024A&A...690A..57S} have a limited potential for this purpose, indirect methods like double-peaked emission lines \citep{2013MNRAS.429.2594B, 2020MNRAS.491.1104D}, periodicities in optical and radio light curves \citep{2012ApJ...759..118B, 2015Natur.518...74G,2022ApJ...926L..35O} and 
cross-symmetric (X, S or Z) radio sources \citep{2007MNRAS.377.1215Z,2017NatAs...1..727K, 2022ApJ...933...98Y} have been employed to short-list dual AGN candidates. 

In spectroscopic surveys using relatively large fibers (e.g., the 3 arcsec diameter fibers used in the Sloan Digital Sky Survey), light from dual AGNs gets integrated/aggregated if the two nuclei are encompassed within the fiber. In that situation, a sufficiently high-resolution optical spectrum may reveal `double-peaked' narrow emission lines (DPNELs), each peak corresponding to the narrow-line region (NLR) of one member of the AGN pair. In the past decades, this method has resulted in spectroscopic selection of hundreds of dual AGN candidates \citep{2009ApJ...705L..76W,2010ApJ...716..866S,2010ApJ...708..427L,2015ApJ...811...14M}. However, such double-peaked emission lines could also arise from other processes, such as bi-conical outflows of warm gas, bulk rotation of the NLR gas, and jet-cloud interactions \citep{2015ApJ...813..103M,2016ApJ...832...67N,2019MNRAS.484.4933R}. In order to further investigate the origin of DPNELs, follow-up observations including high spatial-resolution imaging combined with spatially-resolved spectroscopy are required. In such efforts, only 5\% of the known DPNEL galaxies at $z\le 0.15$ have been followed up with sub-arcsec spatial resolution imaging in optical and NIR bands \citep{2011ApJ...733..103F,2016ApJ...823...50S}. Among them, only about a sixth of such systems having two photometric counterparts within 3'' (less than 10 kpc) have been confirmed with spatially resolved spectroscopy \citep[see,][and references therein]{2021A&A...646A.153S}. Interestingly, out of the Big MAC sample of double-peak AGNs, one parsec-scale binary SMBH, namely J0402+379 \citep{2006ApJ...646...49R}, has robustly been confirmed through the detection of a pair of compact, variable, flat-spectrum, AGN in multi-frequency Very Large Baseline Array (VLBA) observations \citep[][and references therein]{2025ApJS..281...25P}. All this underscores the potential of DPNEL galaxies as the hosts of a dual/binary BH inside the galactic core. 

In addition, X-shaped radio morphologies of jets/lobes have long been linked to the presence of binary SMBHs \citep[e.g.,][]{1980Natur.287..307B} such that gravitational interaction between the two SMBHs can induce jet precession or, ultimately, even jet reorientation (spin-flip) \citep{2002PhDT.......178R, 2002Sci...297.1310M, 2002MNRAS.330..609D, 2013ApJ...768...11B,2023MNRAS.523.1648M,2025MNRAS.536.2025M} in the aftermath of the SMBH coalescence \citep{2001A&A...377...23Z, 2002PhDT.......178R}. For instance, in a multi-frequency VLA observation of X-shaped Radio Galaxy (XRG) J0725+5835, \citet{2022ApJ...933...98Y} found a double radio core with nonthermal radio emission exhibiting the signature of jet reorientation in 5 GHz high-resolution European VLBI Network observations. We also note that several other physical mechanisms, such as diversion of the synchrotron plasma streaming backwards from the terminal hotspots in the two primary lobes, or jet-shell interaction, have also been invoked to explain the formation of XRGs (see reviews by \cite{2012RAA....12..127G} and \cite{2024FrASS..1171101G} and references therein). \cite{2002Sci...297.1310M} have suggested that many Z- or S-shaped radio sources are associated with binary SMBHs approaching coalescence, whereas at least some of the X-shaped radio sources showing off-axis, oppositely-directed radio protrusions may represent merged SMBH systems post the gravitational radiation stage \citep[see also,][]{2001A&A...377...23Z, 2002PhDT.......178R}. In such a merger event, the active SMBH's spin angular momentum axis would flip, and the re-energized jets would normally go off in a significantly changed direction. Thus, binary SMBHs, being the progenitors of the coalesced SMBHs, could eventually be successfully used to constrain the SMBH merger rate and the related physics. On the other hand, wing formation, a marker of X-shaped radio morphology could even begin during the pre-coalescence stage, due to the interaction of the jet pair with the circum-galactic medium of the host elliptical galaxy, which has been set in organized rotation by the in-spiraling of the secondary SMBH (hosted by a smaller galaxy) towards the primary SMBH \citep{2003ApJ...594L.103G}. From all these considerations, it becomes important to try to find and investigate SMBH binaries, their rarity notwithstanding. \par

In this article, we examine the scenario of XRGs hosting dual/binary SMBHs, as suggested by two key indirect indicators: the XRG morphology and DPNELs, while discounting, for the moment, the aforementioned alternative possibilities for the origin of DPNELs. Simultaneously, we also probe the role of radio power output of the AGNs in this context. Some related results based on our Very Long Baseline Interferometry (VLBI) of 4 XRGs are also presented. This paper is structured as follows. Section 2 describes the observations and sample selection. This is followed by the analysis in Section 3 and our results and discussion in Section 4. The conclusions of this study are summarized in Section 5. Throughout, we have assumed a flat Universe with $H_0$ = 70 $\rm km\ s^{-1}\ Mpc^{-1}$, $\Omega_m$ = 0.3, and $\Omega_\Lambda$= 0.7.

\section{Observations and Sample Selection}
To probe the possible binary black hole scenario for the formation of XRGs, we compiled a list of 763 XRGs from the literature. The sources were selected from several published works, including \cite{2007AJ....133.2097C}, \cite{2011ApJS..194...31P}, \cite{2019ApJS..245...17Y}, and \cite{2020ApJS..251....9B}, primarily drawn from the images obtained in the 1.4 GHz VLA FIRST survey \citep{Becker1995}. A summary of the sample, including the catalogs used, the sample sizes (after accounting for any common sources among the catalogs), and details of the final XRG sample, is presented in columns 2 and 3 of Table~\ref{tab:xrg_summary}. 

In particular, the Sample 3 \citep{2019ApJS..245...17Y} contains 290 XRGs, comprising 106 classified as `strong' candidates and 184 as `probable' candidates. Both strong and probable candidates have been included in our sample. Additionally, we have included the 123 new XRGs from \cite{2025ApJ...994...92P}, picked out of the LOFAR Two-metre Sky Survey DR2 \citep{Shimwell2022A&A...659A...1S}, resulting in the final sample of 763 XRGs (see Table~\ref{tab:xrg_summary}).

\begin{table}[ht]
\small
\centering

\caption{Samples of XRG candidates from the literature}
\label{tab:xrg_summary}
\begin{tabular}{lccc}
\hline
\textbf{Source} & \textbf{Sample}& \textbf{Final} & \textbf{Survey} \\
\textbf{} & \textbf{/Overlap} & \textbf{sample} & \\
\hline
1. \cite{2007AJ....133.2097C} & 100/0 & 100 & 1.4 GHz VLA$^{b}$\\
2. \cite{2011ApJS..194...31P} & 156/21& 135 & 1.4 GHz VLA$^{b}$\\
3. \cite{2019ApJS..245...17Y} & 290/25\textsuperscript{a}& 265 & 1.4 GHz VLA$^{b}$ \\
4. \cite{2020ApJS..251....9B} & 161/21& 140 & 1.4 GHz VLA$^{b}$\\
5. \cite{2025ApJ...994...92P}         & 123/0& 123 & 144 MHz LOFAR$^{c}$ \\
\hline
\textbf{Total} & 830  & \textbf{763} \\
\hline
\end{tabular}
\begin{flushleft}
\textsuperscript{a}Includes 106 strong and 184 probable candidates.\\
$^{b}$\citep{Becker1995}\\
$^{c}$ \citep{Shimwell2022A&A...659A...1S}
\end{flushleft}
\end{table}

\begin{figure*}[!htbp]
\centering
\includegraphics[width=0.75\linewidth]{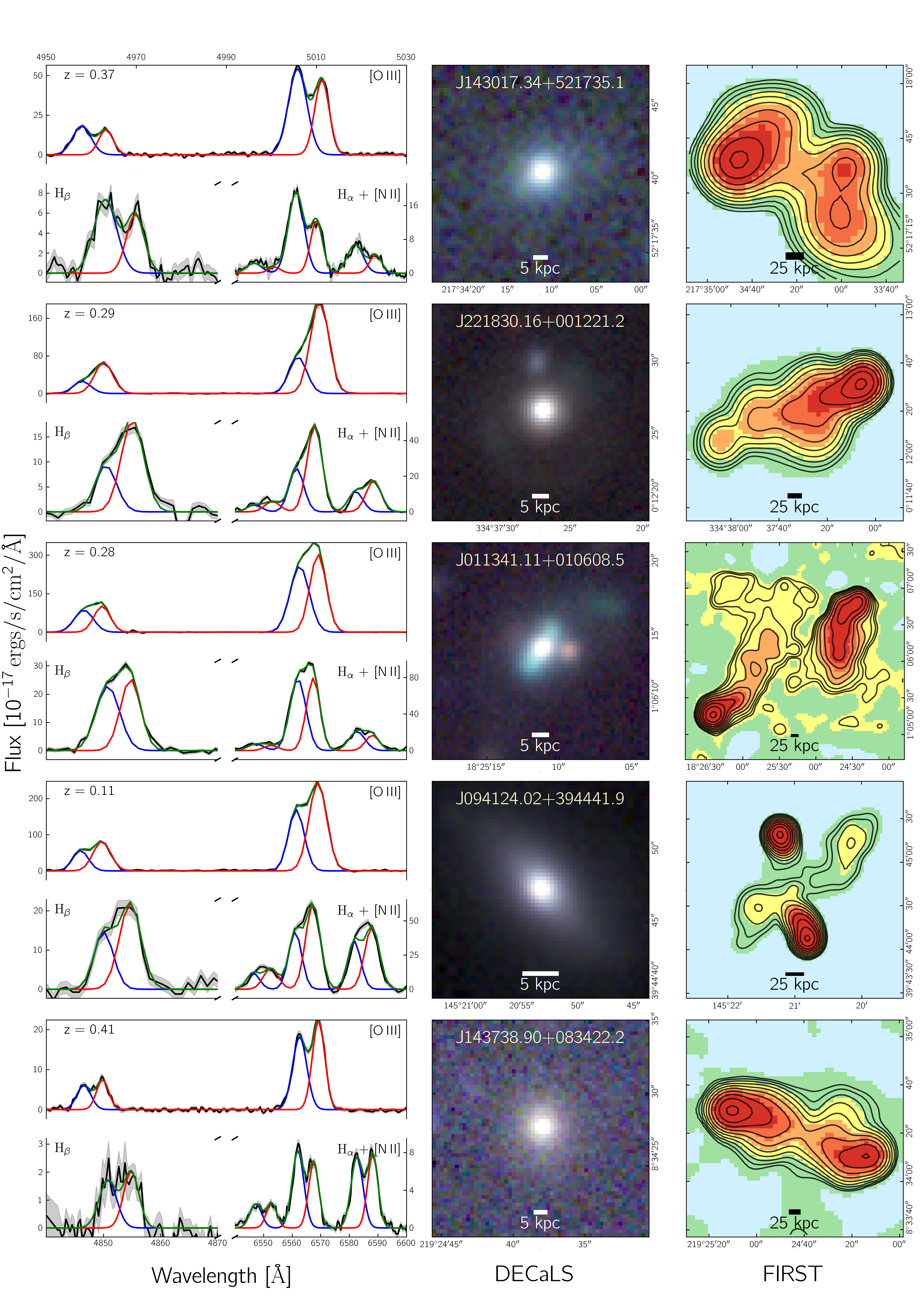}
\caption{\textit{Left}: Double peak emission line profiles for \oiiiab, \hbeta, \halpha, and \niiab\ for five XRGs in the sample. The observed spectra (black) were fitted with Gaussian components, where the blue and red curves represent the blue- and red-ward components, respectively. \textit{Middle}: Color composite (grz) optical images from DECaLS showing the host galaxy morphology. \textit{Right}: Radio continuum contours from the VLA FIRST 1.44 GHz, highlighting the extended radio wings. The 5 kpc (optical) and 25 kpc (radio) scales represent the physical size of the galaxy at the redshift.}
 \label{fig:Collage}
\end{figure*}

\subsection{Optical: Spectral}
\label{sec:sample_selection}
We first searched for the optical spectra in the Sloan Digital Sky Survey (SDSS) DR 18 \citep{2023ApJS..267...44A} and the Dark Energy Spectroscopy Instrument (DESI) Data Release 1 \citep{desicollaboration2025datarelease1dark}. For 487 of the 763 XRGs, spectra could be found, and these span a redshift range,  $0.0054 \le z \le 4.25$. We then primarily used the prominent \oiiib\ nebular emission line to designate a source as a DPNEL galaxy. To characterize the double-peak nature, we estimated the equivalent width of \oiiib\ emission line, and selected a subset of 275 sources for which the \oiiib\ equivalent width is $\ge 3\AA$ \citep{2010MNRAS.403.1036C,2015MNRAS.453.3519P,2017ApJ...846...42K,2025A&A...697A..97Z}(see Section~\ref{lab:analysis}).  Finally, in order to be able to determine the AGN fraction among the selected DPNEL sources, based on the locations of individual sources in the Baldwin, Phillips \& Terlevich (BPT) \citep{Baldwin1981PASP...93....5B} emission-line diagnostic map, a prerequisite was that the \niiab\ and \halpha\ emission lines are covered within the SDSS and DESI spectral wavelength ranges of 3600--10400$\AA$\ and 3600--9800$\AA$, respectively. These requirements imposed together limited our XRG samples to $ z \le 0.58$ and $z \le 0.49$, respectively. Following the afore-mentioned selection criteria, a total of 187 XRGs became available to us for emission-line fitting.

\subsection{VLBA observations}
To address whether the galaxies exhibiting the X-shaped morphology and double-peaked \oiiib\ nebular emission line harbor binary SMBH, we have carried out VLBA observations on a small set of four XRGs that exhibit a prominent double-peaked \oiiib\ nebular emission line. Among the four proposed XRGs from the sample of 187 sources, 3C 223.1 and J1430+5217 were selected from the XRG catalogs of \citet{2007AJ....133.2097C} while TXS1244+200 from \citet{2019ApJS..245...17Y}. In addition, we observed a radio galaxy, J2315+1027 \citep{2025ApJ...994...92P}, with a hint of radio spur as a possible XRG with double-peaked emission lines.

The VLBA 1.5 GHz (L-band) data for these four double-peaked XRGs were acquired using either nine (BJ095A and BJ095D) or ten (BJ095B and BJ095C) antennas of the VLBA in phase-referencing mode, between 08 March 2019 and 03 June 2019 (Proposal ID: BJ095, PI: Joshi, R.). The observations were conducted with an antenna recording rate of 2 Gbps, using two 128-MHz basebands in both left and right circular polarizations, yielding a total recording bandwidth of 512 MHz. The calibrators 3C~286, OQ~208, 3C~48, and 3C~84 were used as fringe finders and bandpass calibrators. The following combos of target - calibrator were used as phase referencing observations: TXS 1244+200 - J1248+2022; J2315+1027 - J2310+1055; 3C~223.1 - J0934+3926; J1437+5112 - J1430+5217. A standard "nodding" phase-referencing strategy was employed, consisting of 4 minutes on the target source followed by 1 minute on the phase calibrator. The observation information can be found in Table \ref{tab1:VLBAobs}.

The observational data from each station were then transferred to the VLBA correlation center in Socorro (New Mexico, USA) for correlation at the Distributed FX (DiFX) correlator \citep[][]{2011PASP..123..275D} with 2s integration time. Then we used the \texttt{AIPS} (Astronomical Image Processing System) software from the US National Radio Astronomy Observatory (NRAO) \citep[][]{2003ASSL..285..109G} to perform data calibration following the standard calibration procedure \footnote{see Appendix C of AIPS Cookbook in http://www.aips.nrao.edu/cook.html}.

The amplitude scale was established using station-specific system temperature measurements and gain curves. We accounted for Earth Orientation Parameters (EOPs) using the U.S. Naval Observatory database, and corrected for ionospheric dispersive delays using Global Positioning System (GPS) derived Total Electron Content (TEC) maps\footnote{\url{https://cddis.nasa.gov}}. Additional corrections for atmospheric opacity and parallactic angle variations were applied using the auxiliary metadata provided with the datasets. Instrumental delays and bandpass responses were calibrated using observations of fringe-finder calibrators within each session. Visibility phases were determined via phase-referencing; specifically, we performed a global fringe-fitting search on the phase-reference calibrator, incorporating its structural model to solve for residual phase delays. These solutions were then interpolated and applied to the target source. The fully calibrated visibilities were exported to Difmap \citep{1997ASPC..125...77S} for imaging and model-fitting. Owing to the moderate flux density of the target source, self-calibration was not implemented.

\begin{table*}
\caption{The VLBA observations}
\label{tab1:VLBAobs}
\centering
\begin{tabular}{cccccc}
\hline
\hline
Code  &  Date & Target: XRG & Phase Calibrator & ToS & Telescopes$^{\rm a}$ \\
   & (YYYY/MM/DD) & (J2000) & (J2000) & (h) & \\
(1) & (2) & (3) & (4) & (5) & (6) \\
\hline
bj095a & 2019/04/27 & TXS~1244$+$200  & J1248$+$2022 & 7.2 & BR,FD,HN,KP,LA,MK,NL,OV,PT \\
bj095b & 2019/03/08 & J2315$+$1027   & J2310$+$1055 & 3.6 & BR,FD,HN,KP,LA,MK,NL,OV,PT,SC \\
bj095c & 2019/03/13 & 3C~223.1     & J0934+3926      & 3.6 & BR,FD,HN,KP,LA,MK,NL,OV,PT,SC \\
bj095d & 2019/03/22 & J1430$+$5217   & J1437$+$5112 & 3.6 & BR,FD,HN,LA,MK,NL,OV,PT,SC \\
\hline
\end{tabular}
\\
\noindent{Note: Col.~1 - Observation code of VLBA sessions; Col.~2 - Observation date; Col.~3 - The target of the observation session; Col.~4 - Phase calibrator of the phase referencing session; 
Col.5~ - Time on the target source of the session;
Col.~6 -  Telescopes participating in the VLBA observation.}
\\
\noindent{$^{\rm a}$VLBA telescopes participating in the observations: BR (Brewster), FD (Fort Davis), HN (Hancock), KP (Kitt Peak), LA (Los Alamos), MK (Mauna Kea), NL (North Liberty), OV (Owens Valley), PT (Pie Town), SC (Saint Croix).} \\
\end{table*}

\section{Analysis \label{sec-3}}
\label{lab:analysis}
In this section, we model the spectra to characterize the double-peak \oiiib\ emission line. At first, we de-reddened the SDSS spectra by applying the galactic dust maps with a Milky Way extinction curve from \cite{1999PASP..111...63F}. We employed the publicly available spectral fitting code PyQSOFit \citep{2018ascl.soft09008G, 2024ApJ...974..153R} only for stellar continuum fitting, which uses two independent sets of eigenspectra --- pure galaxy and pure quasar --- to decompose the spectrum into host galaxy and quasar continuum. The quasar continuum is modeled by a combination of power law and [Fe~{\sc ii}] blend, whereas the Balmer continuum due to the underlying stellar population is modeled with the Penalized Pixel-Fitting (pPXF) routine \citep{Cappellari2023} within PyQSOFit. In brief, the pPXF masks the emission lines and models the spectrum using simple stellar population templates from MILES \citep{2010MNRAS.404.1639V}, which cover a broad metallicity range ($M/H$ from -2.32 to +0.22) and age ranging between 63 Myr to 17 Gyr. 

In order to constrain the double-peak nature of prominent emission lines such as \hbeta, \oiiiab, \halpha, \niiab, and \siiab\ in the XRG spectra, we followed the emission line classification scheme from \cite{2012ApJS..201...31G}. Given that the \oiiib\ is a prominent nebular emission line and an excellent tracer of the extent and kinematics of diffuse ionized gas, we used it as a template for model-fitting all other narrow emission lines of interest, i.e., \hbeta, \halpha, [N~{\sc ii}], and [S~{\sc ii}]. 

We modeled the \oiiib\ emission line using single- and double-component narrow Gaussian profiles, with typical narrow-line widths of $\lesssim 500~\rm km~s^{-1}$. We introduced an additional underlying broad component of width $\gtrsim 2000 \rm km/s$, where required, primarily originating in the broad line region. The fitting has been performed using the non-linear optimization routine \textsc{lmfit} \citep{2021zndo....598352N}. Following \cite{2012ApJS..201...31G} and \cite{2020A&A...641A.171M}, a source is classified as double-peaked in \oiiib\ only if qualifies the following criteria:

\begin{enumerate}
\item The ratio of amplitudes of the two [O{~\sc iii}] Gaussian components is between 1/2.5 and 2.5. This would more likely ensure a clear identification of two peaks and avoid the false positives due to a weak outflow/inflow component. 
\item Both peaks are well separated by at least 3 times the spectral resolution offered by the SDSS (i.e, $\sim$210 $\rm km \, s^{-1}$) and DESI (i.e., $\sim$120 $\rm km \, s^{-1}$).
\item The F-test \citep[Chap. 12.1]{Lupton1993} demonstrates that incorporating the double-peaked profile, as the case may be, yields a significantly better fit compared to the single-peaked model, with a significance level of 5\%. We further verified this by ensuring a lower Bayesian Information Criterion (BIC) value \citep{1978AnSta...6..461S,2007MNRAS.377L..74L} for the double-peaked profile as compared to the single-peaked one. 
\end{enumerate}

An underlying broad component was required to model 27 XRGs. Based on the properties of the \oiiib\ emission line, we classified the sources into four categories: (1) single peak, (2) single peak with a broad underlying component, (3) double peak, and (4) double peak with a broad underlying component. Among the \totxrgs\ XRGs subjected to \oiiib\ line fitting, \oiiidp\ sources were part of 3rd and 4th classification and hence identified as exhibiting double-peaked \oiiib\ emission-line profiles.

The best-fitting \oiiib\ kinematic model for double-peaked sources was subsequently used as a template to fit other prominent narrow emission lines, namely \hbeta, \oiiia, \niiab, \halpha, and \siiab. For these lines, the widths and velocity offsets of the two-fitted Gaussian components were tied to those of the corresponding \oiiiab\ components. The flux ratios of \oiiib/\oiiia\ and \niib/\niia\ were fixed at 3.0 and 2.96, respectively \citep{2006agna.book.....O}. The width of broad component was allowed to vary freely for all emission lines in order to achieve a $\sim$20\% improvement in $\chi^{2}$ wherever necessary.

\begin{figure*}[!htbp]
\centering
\includegraphics[width=0.49\linewidth]{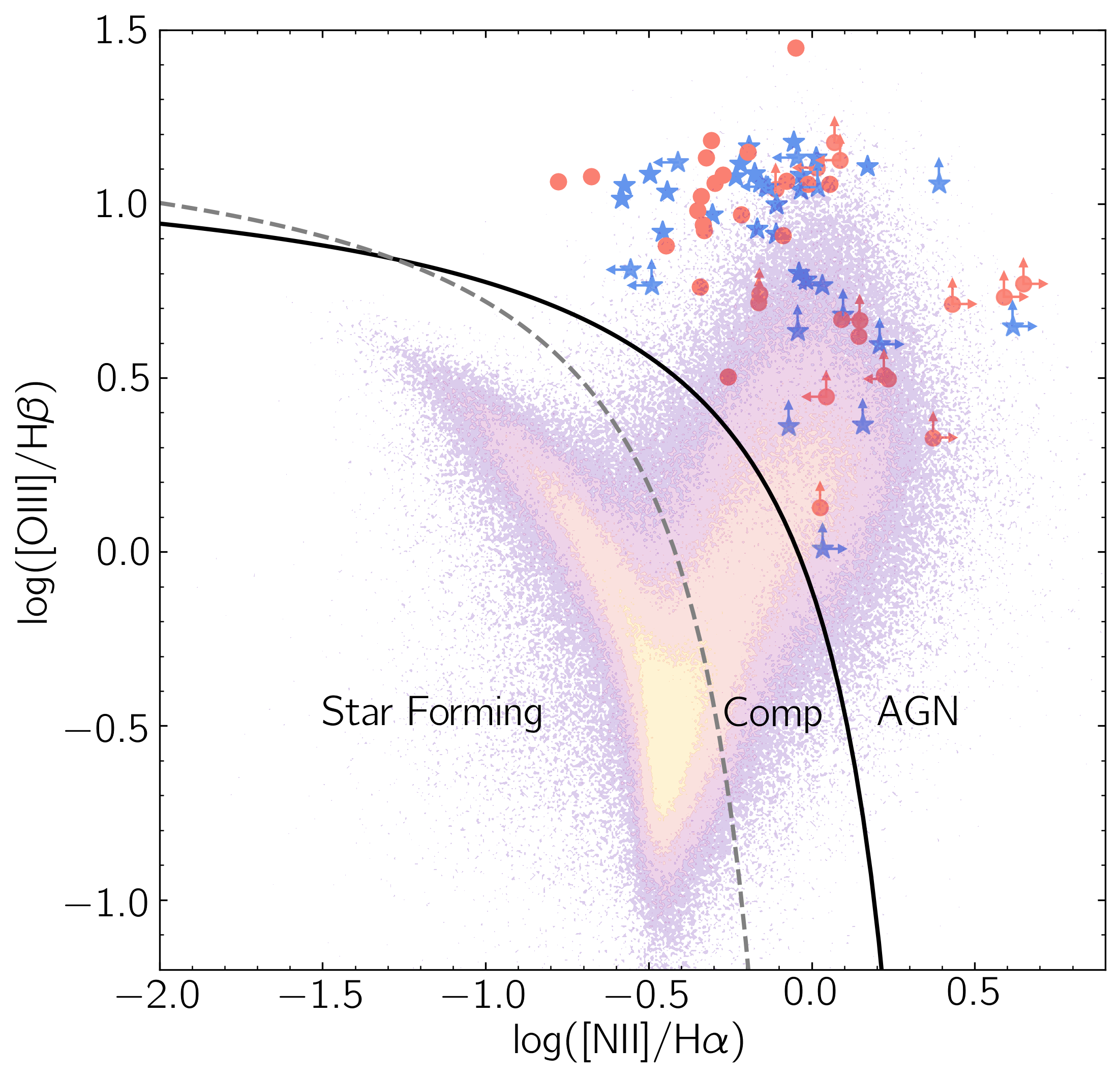}
\includegraphics[width=0.49\linewidth]{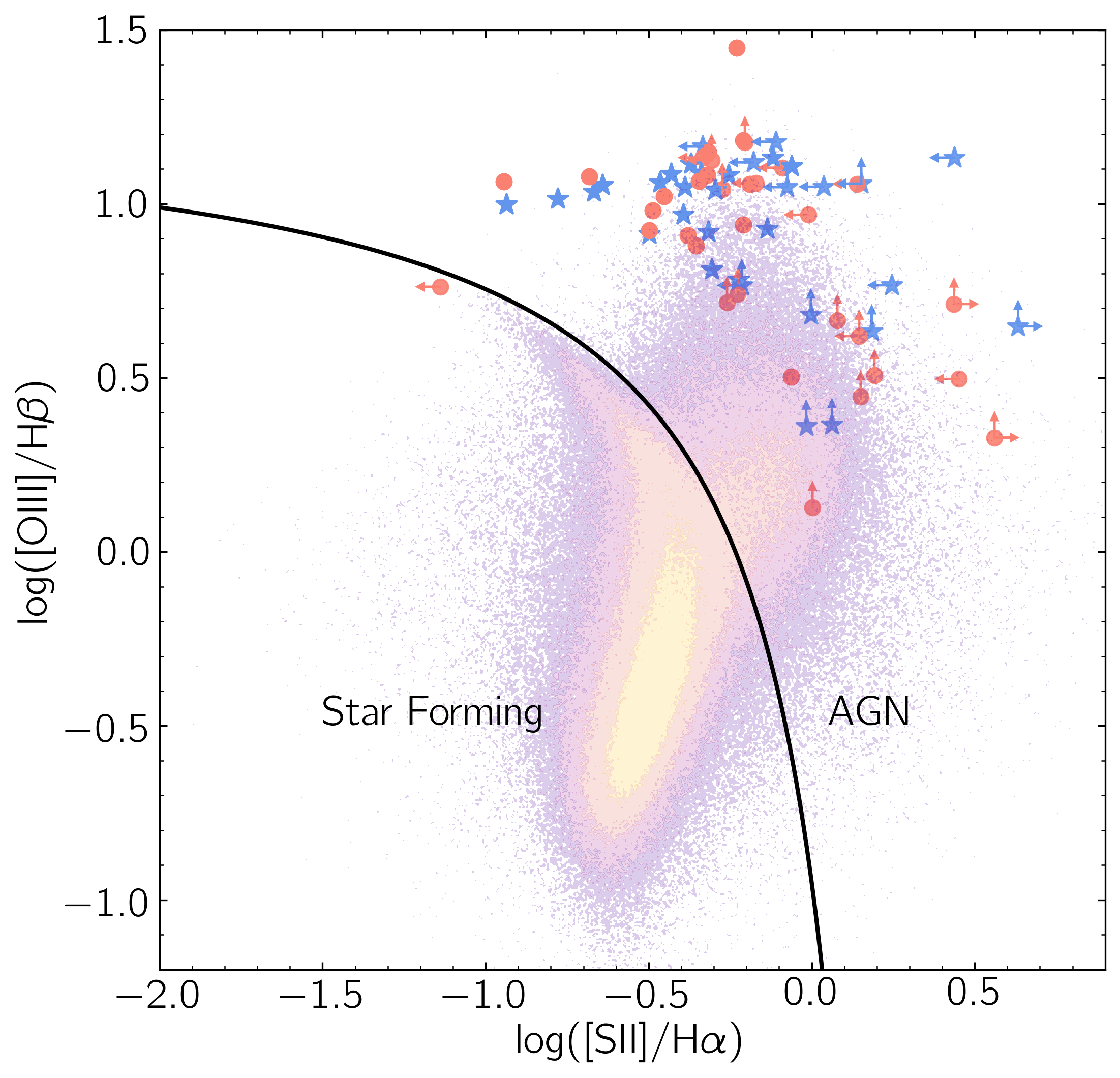}
\caption{ {\it Left Panel:} The [N{~\sc ii}]-BPT diagnostic map for the double-peaked XRGs. The line ratios for the blue and red emission line components are shown as \textit{circle}  and \textit{stars}, respectively. The upper and lower limits are marked with an arrow. The {\it dashed} and {\it solid} line demarcates the AGN from star-forming galaxies based on the BPT classification from \cite{2003MNRAS.346.1055K} and \cite{2006MNRAS.372..961K}, respectively. The region between the two lines is demarcated as {\it Comp}, which stands for composite, indicating photoionization from both star-forming regions and AGN. \textit{Right:} Same as left panel for [S{~\sc ii}]-BPT diagnostic. The colored shaded region in both panels depicts the BPT density map from the SDSS DR8 value-added catalog \citep{2011ApJS..193...29A}.}
 \label{fig:sBPTDiagram}
\end{figure*}

\section{Results and Discussion}
\subsection{Occurrence of dual-AGN candidates in XRGs}
\label{dualagnfrac}
To estimate the fraction of dual-AGN candidates within DPNEL XRGs, we used the BPT diagnostic diagrams, which provide an empirical classification using various emission-line ratios, in order to effectively differentiate between photoionization produced by the harder AGN spectrum (i.e., a larger fraction of higher energy photons) and that produced by hot stars \citep{Baldwin1981PASP...93....5B}. In Figure~\ref{fig:sBPTDiagram}, we show the BPT-diagram based on \niib/\halpha\ versus \oiiib/\hbeta\ and \siiab/\halpha\ versus \oiiib/\hbeta\ line ratios. The extreme starburst line for pure stellar photoionization models by \cite{2003MNRAS.346.1055K} is shown as a {\it solid line} whereas the modified model for dividing the pure star-forming galaxies from AGN and H{\sc~ii} composite sources by \cite{2001ApJ...556..121K} is shown as a {\it dashed line}. We note that, for 3 of our sources, the \halpha\ spectral line region is either corrupted or lies near the edge of the spectral range covered, reducing the sample to 52 XRGs available for BPT analysis. We detected all the nebular emission lines available, i.e., \halpha,~ \hbeta, \niia, at $\ge 3\sigma$ level only for \niibptrobust XRGs.  Additionally, using the upper or lower limits on the line ratios, we could reliably place \niiuplim XRGs in the BPT diagram, yielding a total of 37 out of 52 DPNEL XRGs. The line ratios for the blue and red components are shown as {\it stars} and {\it circles}, whereas the upper limits are marked as arrows, respectively  (see Figure \ref{fig:sBPTDiagram}). We exclude the remaining 15 sources with non-detection of all the lines required to estimate the line ratios.

It is important to note that 33 out of 37 XRGs ($\sim 90\%$) present in the [N{\sc ii}] BPT diagram reside within the locus where the emission lines are primarily excited by the AGN activity. These estimates of fractions include the possibility that the individual data points for which only upper/lower limits are available can accordingly move across the respective diagrams. Similarly, in the [S{\sc ii}] BPT diagram, available for 48 out of 52 XRGs, we detect \siibptrobust galaxies where all emission lines are present at $\ge 3\sigma$. An additional 21 sources can be reliably placed in the BPT diagram based on the upper and lower limits, while the remaining 14 sources with multiple emission lines below $3\sigma$ are discarded from classification. It shows that $\sim 85\%$ (29/34) of these XRGs lie within the AGN domain. Taking both the spectral line diagnostics into account, we find that 35 out of 37 XRGs ($\sim95\%$) exhibit double-peaked components that fall within the AGN locus of the BPT diagram, thus indicating that the line-emitting gas in both kinematic components is predominantly ionized by an AGN. Such an interpretation is consistent with spectroscopic studies of large radio sources, which show that their optical emission-line ratios are primarily governed by AGN photoionization rather than shock-induced ionization  \citep{2000MNRAS.311...23B,2002A&A...383...46M}

\begin{table*}[ht]
\centering
\caption{Summary of model parameters for the \oiiib\ nebular emission in sources with double-peaked [O\,{\sc iii}] profiles. Column (1) lists the SDSS source name, and column (2) gives the spectroscopic redshift. Column (3) provides the black hole mass ($\log M_{\mathrm{BH}}$), while columns (4) and (5) give the logarithm of the NLR radii ($\log R_{\mathrm{NLR,1}}$ and $\log R_{\mathrm{NLR,2}}$), determined using Equation 2 (in Sec. 4.4.4). Columns (6) and (7) list the fluxes of the two [O\,{\sc iii}] components, columns (8) and (9) give their corresponding velocity dispersions, and columns (10) and (11) present the relative redshifts ($z_1$, $z_2$) of the individual components. The final column (12), denoted as $T$, represents the BPT classification: $T=1$ indicates that both components are classified as AGN, $T=0$ corresponds to cases where multiple robust lines are not detected, and classification is not possible, and $T=2$ denotes all other cases such as AGN + star-forming (SF) or SF + SF systems. The table includes the first 10 out of 55 XRGs exhibiting DPNELs$^\ddagger$.
}
\scriptsize
\begin{tabular}{cccccccccccc}
\toprule
\multirow{2}{*}{SDSS Name} & \multirow{2}{*}{$z$} & \multirow{2}{*}{$\log M_\mathrm{BH}$} & \multirow{2}{*}{$\log R_\mathrm{NLR,1}$} & \multirow{2}{*}{$\log R_\mathrm{NLR,2}$} & 
\multicolumn{2}{c}{$F_{\mathrm{[O\,III]}\,5008} \, [\mathrm{erg\,s^{-1}\,cm^{-2}}]$
} & 
\multicolumn{2}{c}{$\sigma_{\mathrm{[OIII]}}$ [km/s]} & 
\multicolumn{2}{c}{$z_{1,2}$} & \multirow{2}{*}{T} \\
\cmidrule(lr){6-7} \cmidrule(lr){8-9} \cmidrule(lr){10-11}
J2000 & & [$\rm M_{\odot}$] & [pc] & [pc] & n1 & n2 & $\sigma_1$ & $\sigma_2$ & $z_1$ & $z_2$ & \\
(1) & (2) & (3) & (4) & (5) & (6) & (7) & (8) & (9) & (10) & (11) &  (12)\\

\midrule
J011341.11+010608.5 & 0.2809 & $8.6 \pm 0.10$ & $3.9 \pm 0.03$ & $3.9 \pm 0.03$ & $1178.96 \pm 11.88$ & $1191.05 \pm 11.42$ & $126.3 \pm 2.4$ & $108.2 \pm 1.6$ & 0.2805 & 0.2813 & 1 \\
J014316.72-011901.0 & 0.5193 & $8.6 \pm 0.10$ & $3.8 \pm 0.03$ & $3.7 \pm 0.03$ & $35.87 \pm 4.76$ & $31.37 \pm 4.74$ & $132.0 \pm 30.5$ & $183.6 \pm 79.3$ & 0.5189 & 0.5199 & 0 \\
J014719.28-085119.6 & 0.4547 & $7.9 \pm 0.12$ & $3.7 \pm 0.03$ & $3.9 \pm 0.03$ & $342.99 \pm 2.92$ & $128.8 \pm 4.2$ & $115.1 \pm 2.2$ & $189.2 \pm 2.8$  & 0.4532 & 0.4550 & 1\\
J021635.79+024400.9 & 0.1451 & $9.3 \pm 0.13$ & $3.3 \pm 0.04$ &  $3.5 \pm 0.04$ & $37.81 \pm 6.31$ & $139.85 \pm 13.45$ & $\geq 42.0$ & $\leq 198.0$ & 0.1445 & 0.1450 & 0 \\
J065514.73+540857.2 & 0.2380 & $9.5 \pm 0.13$ & $3.9 \pm 0.03$ & $4.1 \pm 0.04$ & $809.42 \pm 10.68$ & $4493.35 \pm 14.73$ & $68.8 \pm 2.5$ & $190.1 \pm 1.3$ & 0.2375 & 0.2382 & 1\\
J072423.22+362032.2 & 0.2970 & $9.0 \pm 0.11$ & $3.5 \pm 0.03$ & $3.6 \pm 0.03$ & $44.45 \pm 4.78$ & $68.03 \pm 5.81$ & $92.9 \pm 16.6$ & $109.7 \pm 16.4$ & 0.2965 & 0.2974 & 1 \\
J072645.57+510407.0 & 0.3390 & $10.2 \pm 0.18$ & $4.0 \pm 0.03$ & $3.9 \pm 0.03$ & $1011.98 \pm 8.05$ & $464.51 \pm 6.36$ & $190.2 \pm 7.7$ & $92.3 \pm 4.4$ & 0.3385 & 0.3393 & 1\\
J074125.22+333319.9 & 0.3640 & $8.7 \pm 0.10$ & $4.2 \pm 0.05$ & $4.1 \pm 0.04$ & $1202.66 \pm 9.48$ & $124.57 \pm 5.02$ & $194.7 \pm 5.0$ & $72.8 \pm 8.3$ & 0.3634 & 0.3652 &1\\
J074845.10+232445.8 & 0.1901 & $9.0 \pm 0.11$ & $3.5 \pm 0.03$ & $3.4 \pm 0.04$ & $93.91 \pm 6.13$ & $46.02 \pm 5.24$ & $85.0 \pm 12.8$ & $61.6 \pm 11.6$ & 0.1899 & 0.1905 & 1 \\
J081841.57+150833.5 & 0.3304 & $8.9 \pm 0.11$ & $3.9 \pm 0.03$ & $3.6 \pm 0.03$ & $425.70 \pm 6.68$ & $87.32 \pm 4.82$ & $\leq 170.3$ & $\leq 96.3$ & 0.3299 & 0.3310 & 1 \\
\bottomrule
\end{tabular}
\label{table:xrg}
\begin{flushleft}
$^\ddagger$ The table for remaining sources and fitted flux for other emission lines is provided in a machine-readable format.   
\end{flushleft}
\end{table*}

Here, it is interesting to recall that for the general galaxy population, the fraction of galaxies hosting DPNELs is found to be merely $\sim 1\%$ and only $\sim$ 30\% of these are cases where both line components are classified as AGN, hence dual-AGN candidates \citep{2012ApJS..201...31G, 2020A&A...641A.171M}. In contrast, for the XRGs, we find the fraction of DPNELs to be at least $30\%$ (\oiiidp/\totxrgs) and, moreover, $\sim 95\%$  among them are found to be of dual-AGN type.  To draw reliable conclusions on the higher percentages (or not) of dual AGN in DPNEL XRGs, we design a control sample of double-peaked galaxies from \citet{2012ApJS..201...31G}, closely matched in key physical properties. For each double-peaked XRG in our sample, we searched for a minimum of $\sim$2 unique galaxies which are not picked again, closely matched in stellar-mass range of $\Delta \log(M_{*}) = \pm 0.5$ dex, a redshift of $\Delta z = \pm 0.2$, and a rest-frame color difference of $\Delta (g-r) = \pm 0.3$. We identified at least two control galaxies from SDSS catalog for the 34 DPNEL XRGs. Interestingly, albeit matched in mass and color, only 15/68 ($\sim 22\%$) of the DPNEL galaxy population are classified as dual AGN candidates based on BPT analysis compared to the high fraction of $\sim 95\%$ in the XRGs. This immediately begs the question about the origin of the huge jump in the dual-AGN candidate fraction from $\sim 0.3\%$ for the general galaxy population to $> 30\%$ found here for XRGs. Is this 2 orders-of-magnitude jump linked to the XRG morphology, or more generally,  to the difference in radio luminosity? To investigate this, we now extend the present analysis to encompass classical double radio sources of Fanaroff-Riley type II \citep{Fanaroff1974MNRAS.167P..31F}.

\subsection{Occurrence of dual-AGN candidates at different morphologies}

The above analysis has already provided the occurrence rates of dual-AGN candidates among the general galaxy population and for the XRG population. 
Given the prevalence of overwhelming dual AGN candidate signatures in DPNEL XRGs, it is worth exploring the AGN fraction across different radio morphologies, including DPNEL FR-II radio galaxies whose radio powers can be comparable to or even exceed those of XRGs. For this, we employ the FR-II radio source catalog from \citet{2024RAA....24c5021L}, comprising 45,241 FR-II galaxies. It spans a wide range of luminosities $\mathrm{2.63 \times 10^{22} \le L^{1.4GHz}_{rad} \le 6.76 \times 10^{29} \, W \, Hz^{-1}}$ and redshift $0.0 \le z \le 5.01$. We further searched for the optical spectra in SDSS and DESI datasets, following the selection criteria outlined in  Sec:~\ref{sec:sample_selection}, resulting in 199 sources with \oiiib\ emission line equivalent width of $\ge 3 \, \AA$. By employing the multiple emission line modeling, explained in Sec:~\ref{lab:analysis}, we found 21 ($\sim 10\%$) FR-II sources to have double-peaked emission lines. For 16 of these sources with the corresponding line ratios determined within the above-specified limits, the BPT line diagnostic classified all but one as dual AGN candidates, yielding a $\sim 94\%$ dual AGN candidate fraction in FR-II radio galaxies. We note that the DPNEL XRGs and FR-II galaxy samples share similar characteristics,  spanning similar radio luminosities in the range of $\rm 2.5 \times 10^{23} \le L^{1.4GHz}_{rad} \le 7.64 \times 10^{26} \, W \, Hz^{-1}$ with median luminosities of $\rm L^{1.4GHz}_{rad} = 1.2 \times 10^{26} \, W \, Hz^{-1}$ and $\rm 0.5 \times 10^{26} \, W \, Hz^{-1}$, respectively. In addition, both samples have similar average stellar mass of  $\rm log(M_*/M_{\odot}) = 10.9 \pm 0.3$, and $\rm  11.1 \pm 0.3$, respectively.
The high dual AGN fractions observed in both FR-II and XRG populations indicate that the occurrence of dual AGN may not be primarily driven by radio morphology, and could instead be associated with radio luminosity.

\subsection{Occurrence of dual-AGN candidates at different radio luminosities}

\begin{figure}[!htbp]
\centering
\includegraphics[width=\linewidth]{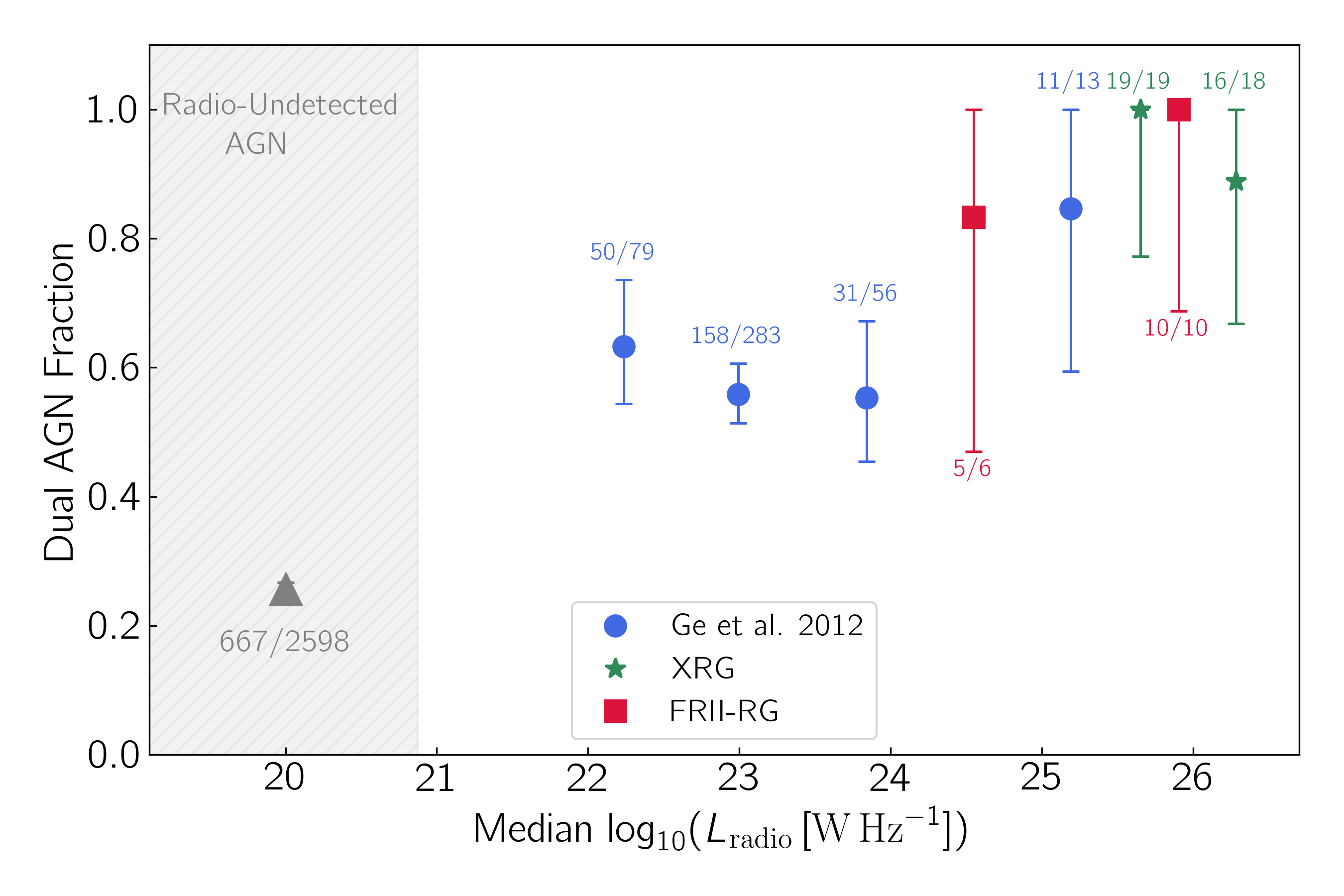}
\caption{The radio luminosity dependence of dual AGN fraction in double-peaked emission line galaxies based on  BPT diagnostics. The radio-undetected and radio-detected DPNEL galaxies from \cite{2012ApJS..201...31G} are shown as triangles and circles, respectively. The errors correspond to 1$\sigma$ confidence level and are computed using the standard Poisson statistics \cite{1986ApJ...303..336G}.  The higher fraction of dual AGNs in both XRGs ({\it stars}) and FR-II radio galaxies ({\it squares}) underscores the radio dependence. }
 \label{fig:fracvsradiolum}
\end{figure}

In order to examine the radio luminosity dependence of the dual AGN fraction in DPNEL galaxies, we matched the 3030 DPNEL galaxies from \citet{2012ApJS..201...31G} against the 1.4 GHz FIRST catalog \citep{Becker1995}, resulting in a subsample of 431 radio-detected DPNEL galaxies. In Figure~\ref{fig:fracvsradiolum}, we plot the dual AGN fractions for radio-detected DPNEL galaxies, which include both the XRGs and FR-II sources. It is evident that the dual AGN fraction strongly correlates with the radio luminosity. While among the radio-detected DPNEL galaxies, overall $\sim 58\%$ are classified as dual AGN, the fraction crosses 94\% for the XRGs and FR-II radio galaxies. Interestingly, the dual AGN fraction drops to 25\% in the case of radio-undetected DPNEL galaxies (see Fig:~\ref{fig:fracvsradiolum}). Further, to make a robust comparison, we select a control sample of DPNEL galaxies with radio luminosities outside the range of our parent sample of 55 DPNEL XRGs, matched on all other physical properties (redshift, color, and stellar mass), and filter 369 sources out of the 431 radio-detected DPNEL galaxies. It is worth noting that 199/369 ($\sim 54\%$) of them have been classified as dual AGN in the BPT diagnostics. This reiterates the higher detection fraction of DPNELs in radio-bright galaxies.

\subsection{XRGs with and without double-peak emission line profiles}

Here, we compare the physical properties of XRGs with and without DPNELs, along with FRI and FRII-type radio galaxies. Recall that the mass of the central SMBHs is found to be tightly correlated with the stellar mass of the bulge of the host galaxy \citep{Kormendy2013ARA&A..51..511K}. We determine the SMBH mass based on the galaxy stellar mass versus black hole mass scaling relation from \citet[][see below]{2020ARA&A..58..257G} for parent sample of \totxrgs XRGs, including \oiiidp XRGs with DPNEL. For 141 XRGs with SDSS spectra, we obtained the multi-band optical (\textit{u, g, r, i, z}) photometry data from the SDSS DR16 survey \citep{2020ApJS..249....3A}, complemented by mid-infrared (3.4 and 4.6 $\mu$m; W1 and W2 bands) data from the unWISE catalog \citep{2019ApJS..240...30S}. Next, we model the spectral energy distribution (SED) of 141 XRGs using the piXedfit routine, which uses a Bayesian technique to estimate the underlying parameters \citep{2021ApJS..254...15A}. For this, we used the model with a wide parameter space, including double power-law star formation history~\citep{2018MNRAS.480.4379C}, initial mass function \citep{Chabrier2003PASP..115..763C},  dust attenuation ($0 \leq \rm{Av} \leq 3$) \citep{2000ApJ...539..718C}, and dust emission \citep{2007ApJ...657..810D}. Further, to account for the contribution from the active black hole, we added the emission from the dust torus heated by the AGN \citep{2008ApJ...685..147N,2008ApJ...685..160N}. In addition, for 46 DESI XRGs, we use the stellar mass estimates provided by the DESI value-added catalog~\citep{2024A&A...691A.308S}, where the SED modelling is performed with a similar parameter space using DESI \textit{g, r, z} and W1 and W2 bands. For all but 3 XRGs with the best fit SED, the stellar masses range from ${9.83} \le \rm log (M_{\star}/M_{\odot}) \le {12.46}$, with a median value of ${10.86}$.  Using the $M_{BH} - M_{*}$ scaling relation by \citet[][see their Table 9]{2020ARA&A..58..257G}, we estimate the black hole mass as:
\begin{equation}
\label{eqn:bbhmas}
\begin{split}
    \log M_{BH} = \, & (7.89 \, \pm \, 0.09) + (1.33 \, \pm \, 0.12)\log\left(\frac{M_*}{3 \times 10^{10} M_{\odot}} \right) \\
    & + (0.65 \, \pm \, 0.05).
\end{split}
\end{equation}

\begin{figure}[!htbp]
\centering
\includegraphics[width=\linewidth]{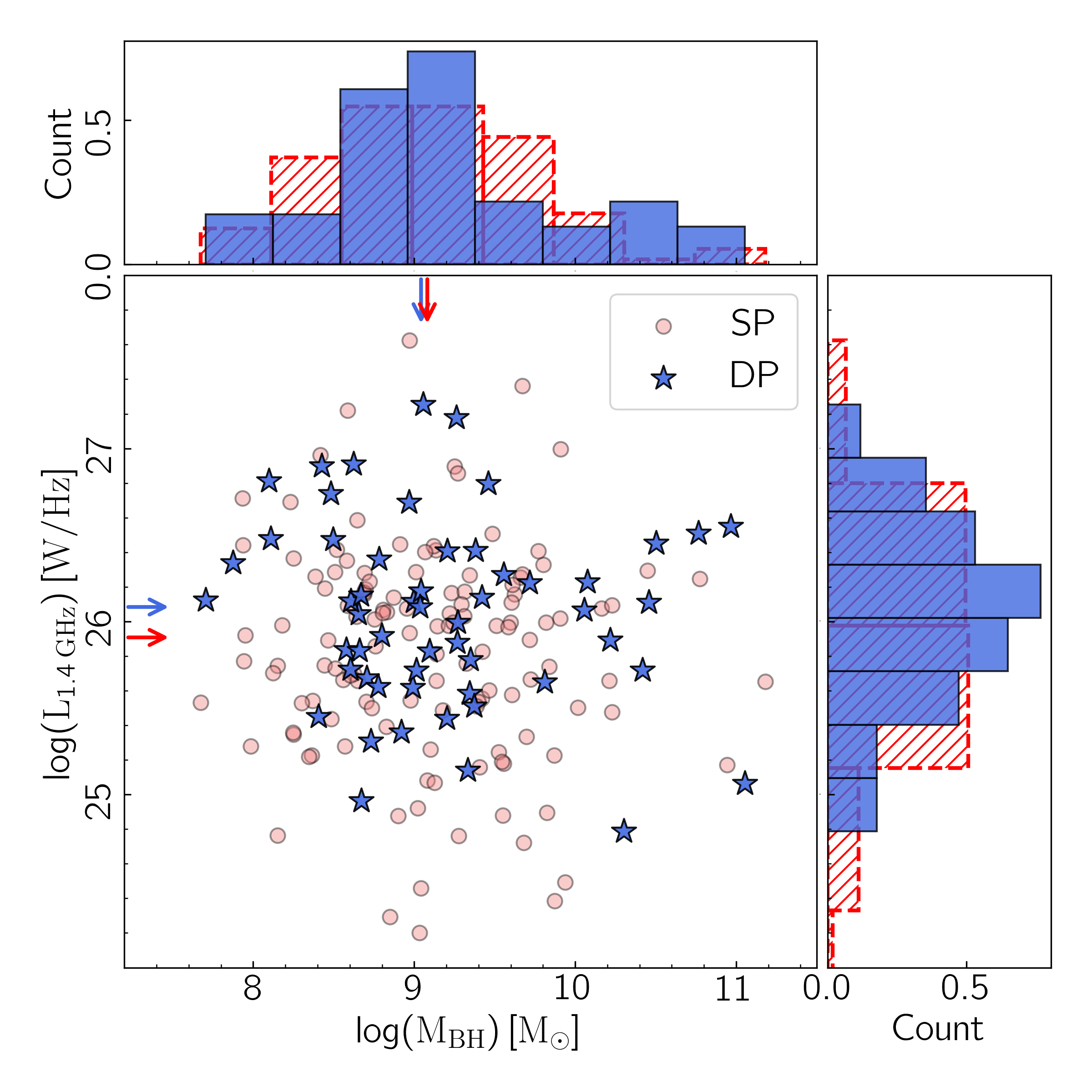}
  \caption{The black hole mass versus 1.4 GHz radio luminosity for the single-peaked ({\it circle}) and double-peaked ({\it stars}) XRGs. The double-peaked XRGs ($\langle \rm log\ M_{BH} \rangle$ = \dpbhmassmean $ \rm \pm \ \dpbhmassmeanerr $ $M_{\odot}$) host very similar black hole masses as compared to single-peaked XRGs ($\langle \rm log\ M_{BH}  \rangle$ = \spbhmassmean $\ \pm \ \spbhmassmeanerr \ M_{\odot}$). The radio luminosities of double-peaked XRGs ($\rm \langle \log L_{1.4 \ GHz} \rangle = \dpradiolummean \pm \dpradiolummeanerr \ W \ Hz^{-1}$) are also similar to those of single-peaked XRGs ($\rm \langle \log L_{1.4 \ GHz} \rangle = \spradiolummean \pm \spradiolummeanerr \ W \ Hz^{-1}$).  }
     \label{fig:BHvsLRAD}
\end{figure}

The estimated black hole masses for the XRGs with DPNELs are presented in column 2 of Table~\ref{table:xrg}. In the upper panel of Figure~\ref{fig:BHvsLRAD}, we compare the SMBH mass distributions of XRGs with and without DPNELs.

The estimated black hole masses for XRGs with DPNEL range from  $\dpbhmasslow  \le \rm log\ M_{BH}\ [M_{\odot}] \le \dpbhmassup$, with a median of  $\langle \rm log\ M_{BH} [M_{\odot}] \rangle$ = \dpbhmassmean $ \rm \pm \ \dpbhmassmeanerr $. The XRGs without DPNELs exhibit a similar average black hole mass of $\langle \rm log\ M_{BH} \ [M_{\odot}] \rangle$ = \spbhmassmean $\ \pm \ \spbhmassmeanerr $, ranging from  $\spbhmasslow  \le \rm log\ M_{\odot}\ [M_{\odot}] \le \spbhmassup$. A Kolmogorov–Smirnov (KS) test yields a null probability value of $p_{\text{null}} = 0.58$, supporting the null hypothesis that the SMBH mass distributions of XRGs with and without DPNELs are drawn from the same parent population. This suggests that XRGs with and without double-peaked emission lines have undergone a similar evolutionary phase. All the DPNEL XRGs showing two active AGNs in a BPT diagram indicate that the XRGs are likely formed through mergers. Thus, the similar black hole masses in XRGs with and without DPNEL can be reconciled under a common evolutionary phase. A larger sample of double-peak XRGs from next-generation optical and radio surveys should help further understand this.

In previous efforts to constrain the formation scenarios of XRGs, involving either the merger of two SMBHs leading to the jet reorientation or the presence of two active SMBHs, \citet{2011A&A...527A..38M} analyzed 29 XRGs from a sample of 100 sources selected by \cite{2007AJ....133.2097C}. They found that the XRGs are hosted in elliptical galaxies, likely formed through hierarchical mergers  \citep[see also,][]{1983FCPh....9....1E, 1988ApJ...327..507F, 2006ApJ...652..864H}. Additionally, the average black hole mass in XRGs ($\rm  \langle \log M_{BH} [M_{\odot}] \rangle = 8.32 \rm$) was found to be 1.94 times higher than that of a limited control sample of 20 radio galaxies, including 10 each of non FRII ellipticals and FRII radio galaxies,  with similar redshifts and comparable optical and radio luminosities, i.e.,  $\rm  \langle \log M_{BH} [ M_{\odot}] \rangle = 8.04 \rm$. However, using a larger set of 106 XRGs,  \cite{2019ApJ...887..266J} have found the average black hole mass of XRGs ($\rm  \langle \log M_{BH} [M_{\odot}] \rangle = 8.81 \rm $) to be slightly lower than that of a control sample of 388 FRII radio sources hosting powerful jets ($\rm  \langle \log M_{BH} \rangle = 9.07 [M_{\odot}] \rm$). The relatively lower black hole mass of XRGs compared to their more powerful FRII counterparts suggests that even minor mergers can cause jet reorientation, contributing to their distinctive X-shaped radio morphology.

In Figure~\ref{fig:BHvsLRAD}, we compare the black hole mass versus the radio luminosity of XRGs with ({\it star}) and without ({\it circle}) double peaks. For this, we estimate the radio flux using the NRAO VLA Sky Survey (NVSS) at 1.4 GHz, which uses a much larger beam of $45^{\prime\prime}$ diameter. The NVSS survey, with good UV coverage, is known to be sensitive to diffuse radio emission extending up to at least 10 arcmin \citep{Condon_nvss_1998AJ....115.1693C}. Assuming a spectral index of $\alpha \sim -$0.7, we estimate the radio luminosity for all of the XRGs. The radio luminosity for XRGs with DPNEL range from \dpradiolumlow $\le \rm \log L_{1.4 \ GHz} \ [W/Hz] \le \dpradiolumup$, with an average of $\langle \rm \log L_{1.4 \ GHz} \rangle = \dpradiolummean \ W \ Hz^{-1}$. We also see that XRGs without DPNEL show similar radio luminosities, ranging from $\spradiolumlow \le \rm \log L_{1.4 \ GHz} \ [W/Hz] \le \spradiolumup$ with average luminosity of $\langle \rm \log L_{1.4 \ GHz} \rangle = \spradiolummean \ W \ Hz^{-1}$, hinting at a common powering source. This aligns with the findings of \cite{2019ApJS..245...17Y} that XRGs have intermediate radio luminosity between FRII and FRI radio sources. However, we report no correlation between black hole mass and radio luminosity in our parent sample of XRGs, with a Spearman's rank correlation coefficient of $r_s = -0.061$ and a null probability of $p_{null}$ = 0.41.

\begin{figure}[!htbp]
\centering

\includegraphics[
width=\linewidth,
height=0.26\textheight,
keepaspectratio
]{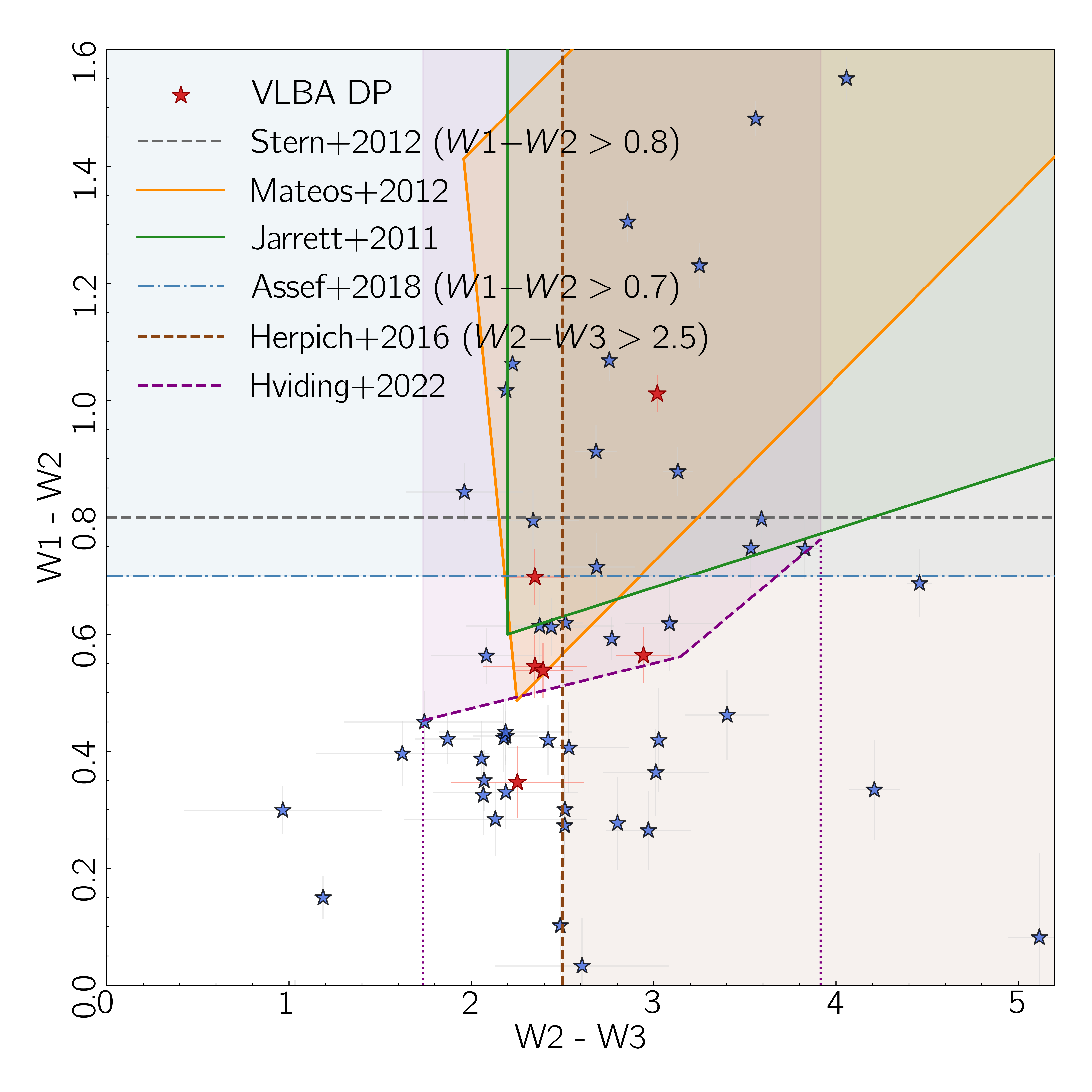}

\vspace{0.1cm}

\includegraphics[
width=\linewidth,
height=0.26\textheight,
keepaspectratio
]{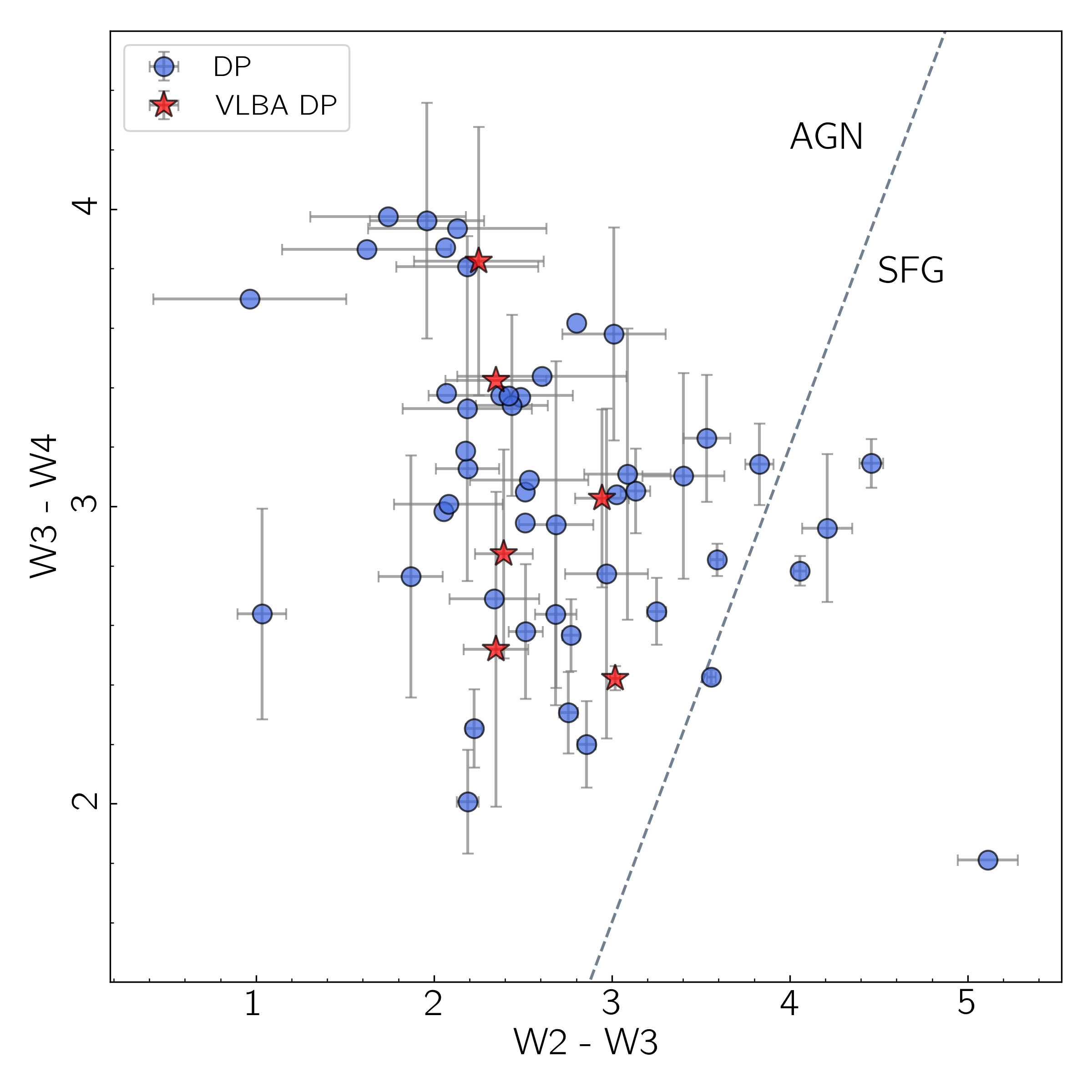}

\vspace{0.1cm}

\includegraphics[
width=\linewidth,
height=0.26\textheight,
keepaspectratio
]{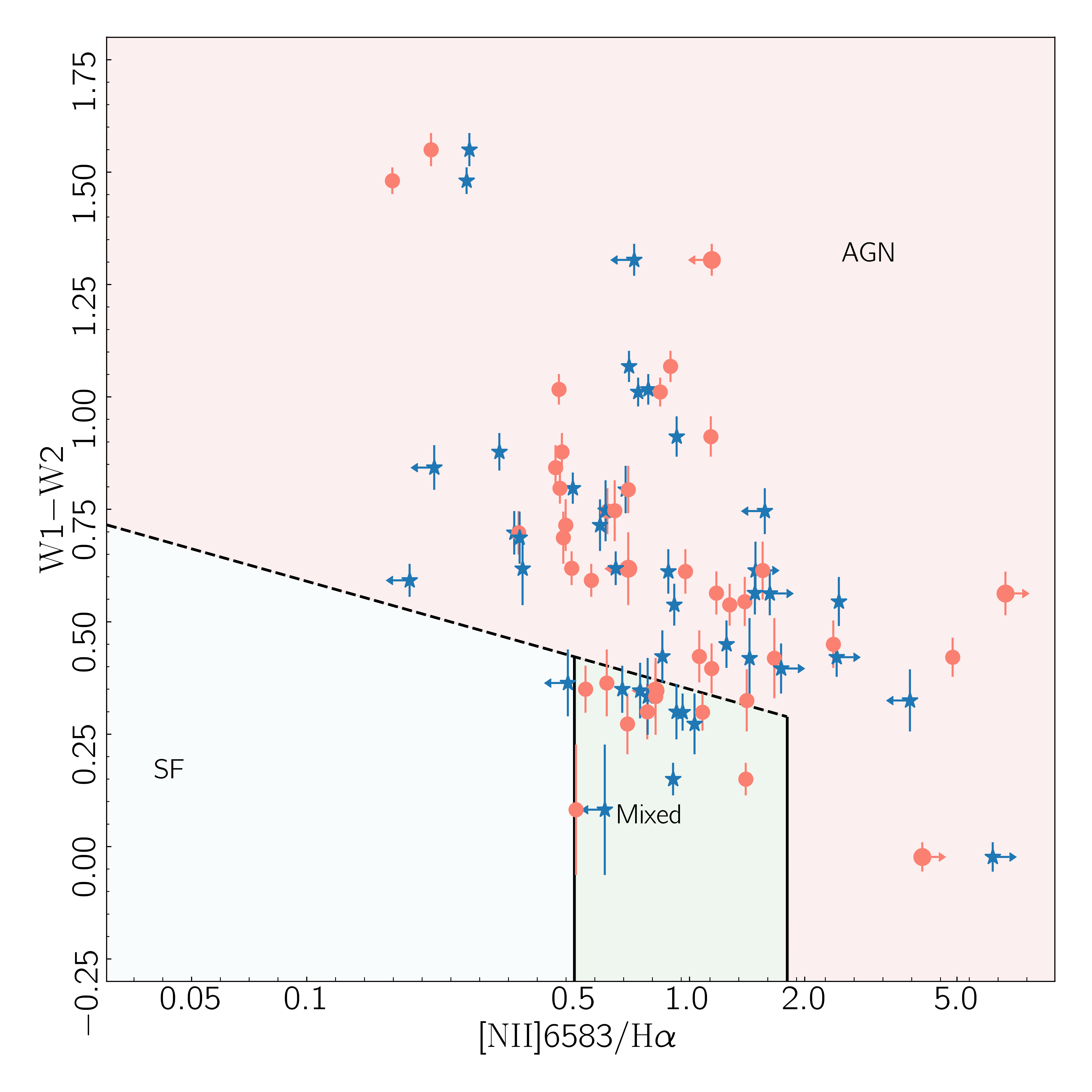}

\caption{
The mid-infrared color diagnostic diagram for the XRGs hosting DPNELs ({\it circle}) with emphasis on the VLBA sample ({\it star}). In the top panel, 69 \% of the sample is classified to host AGN following at least one of the diagnostic criteria. While in the middle panel, the {\it dotted} line \citep{2015AJ....149..192C} separates the AGN locus from the star-forming region. All but 5 (91\%) indicate AGN photoionization in agreement with the BPT diagnostic (Figure~\ref{fig:sBPTDiagram}). In the bottom panel, \cite{2020ApJ...903...91Y} classification scheme indicates 79 \% of the sources to have both components classified as AGNs.
}
\label{fig:W1W2}
\end{figure}

Finally, we explore the mid-infrared (MIR) color properties of the double-peaked XRGs. The MIR diagnostic colors can be used to distinguish star-forming galaxies from AGNs even in cases where one or both of the AGNs are obscured \citep{2004ApJS..154..166L, 2005ApJ...631..163S, 2015AJ....149..192C}. Among the 55 XRGs harboring DPNELs, we could obtain the colors of 54 XRGs using four mid-infrared bands of the Wide-field Infrared Survey Explorer (WISE) survey  \citep{2010AJ....140.1868W} W1 (3.4$\mu$m), W2 (4.6$\mu$m), W3 (12$\mu$m), and W4 (22$\mu$m) in Vega magnitudes. We use three different classification schemes, a) $W1 - W2$ vs $W2 - W3$ \citep{2012ApJ...753...30S, 2012MNRAS.426.3271M, 2011AAS...21832810J, 2018ApJS..234...23A, 2016MNRAS.462.1826H, 2022AJ....163..224H}, b) $W3 - W4$ vs $W2 - W3$ \citep[see Section 3.1 of][]{2015AJ....149..192C} and c) $W1 - W2$ vs \niib/\halpha  \citep{2020ApJ...903...91Y}. In $W1 - W2$ vs $W2 - W3$ WISE classifications, we find that 38/55 (69 \%) are ionized by AGNs following at least one criterion. Similarly, in the $W3 - W4$ vs $W2 - W3$ classification, we find that 49/54 (91\%) of the sources are ionized by AGNs. Finally, $W1 - W2$ vs \niib/\halpha classifies 32 out of 41 (78 \%) with detected \niib/\halpha \, as candidate dual AGNs. The high fraction in these classifications is consistent with that determined from the BPT diagrams (See Section: \ref{dualagnfrac}).

\subsection{Parsec-scale view of XRGs}

\begin{table*}[ht]
\centering
\caption{Summary of VLBA observation details for the dual AGN candidates. Columns: (1) source name; (2) observation date; (3) central observing frequency ($\nu_{\rm cen}$) in GHz; (4) synthesized beam size in milliarcseconds (mas); (5) potential separation that could be resolved at the given frequency in parsecs (pc); (6) position angle (PA) of the separation vector in degrees; (7) image $\rm rms$ noise in mJy; (8) spectroscopic redshift ($z$); (9) spectral index between the indicated frequency pairs; (10) Gaia-VLBA positional offsets in parsec. Multiple rows per source correspond to observations at different frequencies. References and notes for individual sources are provided below the table.}

\begin{tabular}{cccccccccc}
\hline
\hline

  \multicolumn{1}{c}{Source} &
  \multicolumn{1}{c}{Obs Date} &
  \multicolumn{1}{c}{$\rm \nu_{cen}$} &
  \multicolumn{1}{c}{Beam} &
  \multicolumn{1}{c}{Separation} &
  \multicolumn{1}{c}{PA} &
  \multicolumn{1}{c}{$\rm rms$} &
  \multicolumn{1}{c}{z} &
  \multicolumn{1}{c}{Spectral Index} &
  \multicolumn{1}{c}{VLBA-Gaia offset}\\

        & & (GHz) & (mas) & (pc) & (deg) & (mJy) & & & (pc)\\
(1) & (2) & (3) & (4) & (5) & (6) & (7) & (8) & (9) & (10) \\
\hline
3C 223.1 & 2019-03-13 & 1.55 & $10.10 \times 4.34$ & 20 & -0.74 & 0.02 & 0.11 & 0.25 & 22\\
    & 2018-08-11\textsuperscript{a} & 4.98 & $3.46 \times 1.62$ & 6.7 & 15.56 & 0.02 &  & 0.32$^\alpha$ &\\
J1115$+$4314$^{\ddagger}$ & 2016-08-14 & 4.34 & $4.73 \times 1.89$ & 29 & 37.4 & 0.07 & 0.46 & 0.54$^\beta$ & 74\\
                        & 2016-08-14 & 7.62 & $2.90 \times 1.07$ & 18 & 39.88 & 0.09 & & &\\
J1134$+$3835$^{\ddagger}$ & 2016-08-19 & 4.34 & $5.11 \times 4.34$ & 30 & 36.73 & 0.08 & 0.50 & 0.23 & \\
                        & 2016-08-19 & 7.62 & $3.05 \times 1.30$ & 18 & 6.84 & 0.09 & & & \\
TXS 1244$+$200 & 2019-04-27 & 1.55 & $16.01 \times 5.70$ & 82 & -15.80 & 0.01 & 0.43 & 0.24 &\\
J1430$+$5217 & 2019-03-22 & 1.55 & $9.79 \times 3.78$ & 61 & -4.70 & 0.02 & 0.37 & 0.00 & 8\\
           & 2022-05-25 & 4.35 & $5.89 \times 3.4$ & 30 & 32.8 & 0.10 & & 0.19$^\gamma$ &\\
           & 2022-05-25 & 7.62 & $7.66 \times 2.27$ & 39 & 87.12 & 0.20 & & &\\
J1604$+$3221$^{\dagger}$ & 2014-01-27 & 4.34 & $4.61 \times 2.03$ & 27 & 21.40 & 0.09 & 0.45 & 0.01$^\delta$ &128\\
                       & 2014-01-27 & 7.62 & $2.65 \times 1.25$ & 15 & 22.007 & 0.10 & & &\\

J2315$+$1027 & 2019-03-08 & 1.55 & $11.80 \times 4.69$ & 42 & -4.72 & 0.02 & 0.26 & 0.49 & 137\\
\hline
\end{tabular}
\begin{flushleft}
$^{\ddagger}$\citet{2020ApJS..251....9B}\\
$^{\dagger}$\cite{2025ApJ...994...92P}\\
\textsuperscript{a}\cite{2024MNRAS.530.4902S} \\
\textsuperscript{$\alpha$}non-simultaneous spectral index between 1.55\,GHz (2019-03-13) and 4.98\,GHz (2018-08-11)\\
\textsuperscript{$\beta$}between 4.34\,GHz and 7.62\,GHz on 2016-08-14\\
\textsuperscript{$\gamma$}between 4.34\,GHz and 7.62\,GHz on 2022-05-25\\
\textsuperscript{$\delta$}between 4.34\,GHz and 7.62\,GHz on 2014-01-27\\

\end{flushleft}
\label{table:vlbaxrg}
\end{table*}

In previous studies, galaxies hosting DPNELs have been proven to be potential dual AGN candidates, and their binary structures have been explored using high-resolution VLA observations \citep{2015ApJ...799...72F, 2015ApJ...815L...6F, 2023ApJ...945...73G}. However, VLBA, with its very high angular resolution, probes the radio properties at parsec scales, which are crucial for studying the origins of radio emission \citep{2024ApJ...969...36X}. Notably, \cite{2022ApJ...933...98Y} conducts a high-resolution 5 GHz VLBA observation of XRG J0725+5835, which has a resemblance in radio morphology to the archetypal binary AGN 0402+379 \citep{2006ApJ...646...49R}. In this observation, two milliarcsecond-scale non-thermal radio cores were detected. These cores coincide with two optical counterparts with similar photometric redshifts and magnitudes, corresponding to a dual AGN. Similarly, \citet{2024MNRAS.530.4902S} has recently examined the reliability of identifying dual AGNs in a sample of six sources exhibiting both X-shaped radio morphology and DPNELs using VLBA observations and found 3 XRGs with two resolved central components, making them strong candidates for binary SMBHs.

In what follows, we discuss the 1.4 GHz VLBA maps from our VLBA campaign of four XRGs exhibiting double-peak emission lines (see, Figure 6). We also searched for the 55 double-peaked XRGs in the ``Radio Fundamental Catalogue" \citep{2025yCat..22760038P}, which includes all the sources observed with VLBA for absolute astrometry and geodesy from April 11, 1980, to September 5, 2024. As a result, we identified three additional sources with VLBA maps,  J1115+4314, J1134+3835, and J1604+3221, which are reproduced in Fig.\ 6. We also found additional 4.4 GHz and 7.6 GHz observations to supplement our 1.4 GHz observations of J1430+5217. The sample summary, including source name, date of observation, frequency, beam size, $\mathrm{rms}$ noise, redshift, separation between the optical and radio cores, and in-band spectral index (or dual-band in 'Spectral Index' column), is listed in Table~\ref{table:vlbaxrg}.

\begin{figure*}[!htbp]
\centering
\includegraphics[width=.49\linewidth]{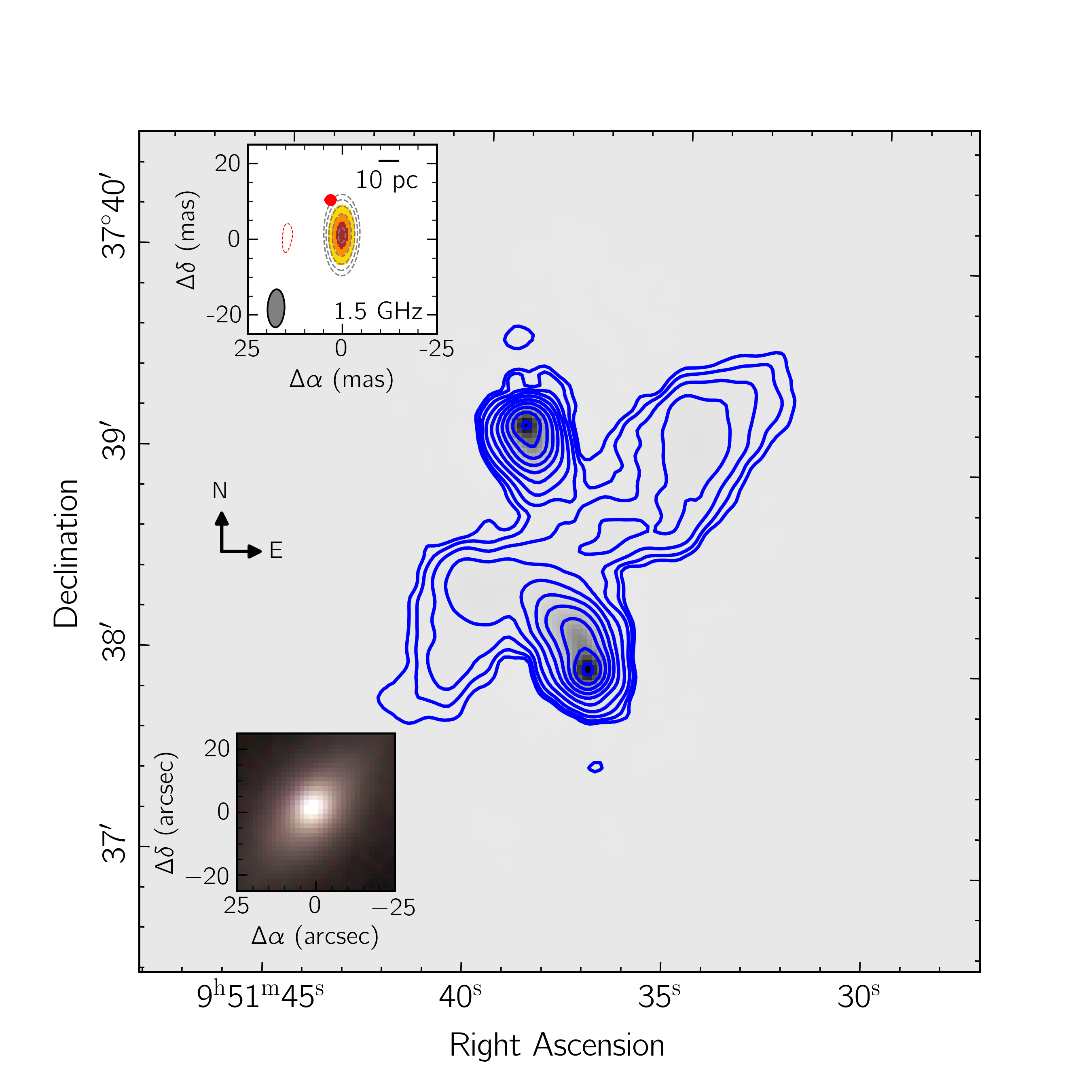}
\includegraphics[width=.49\linewidth]{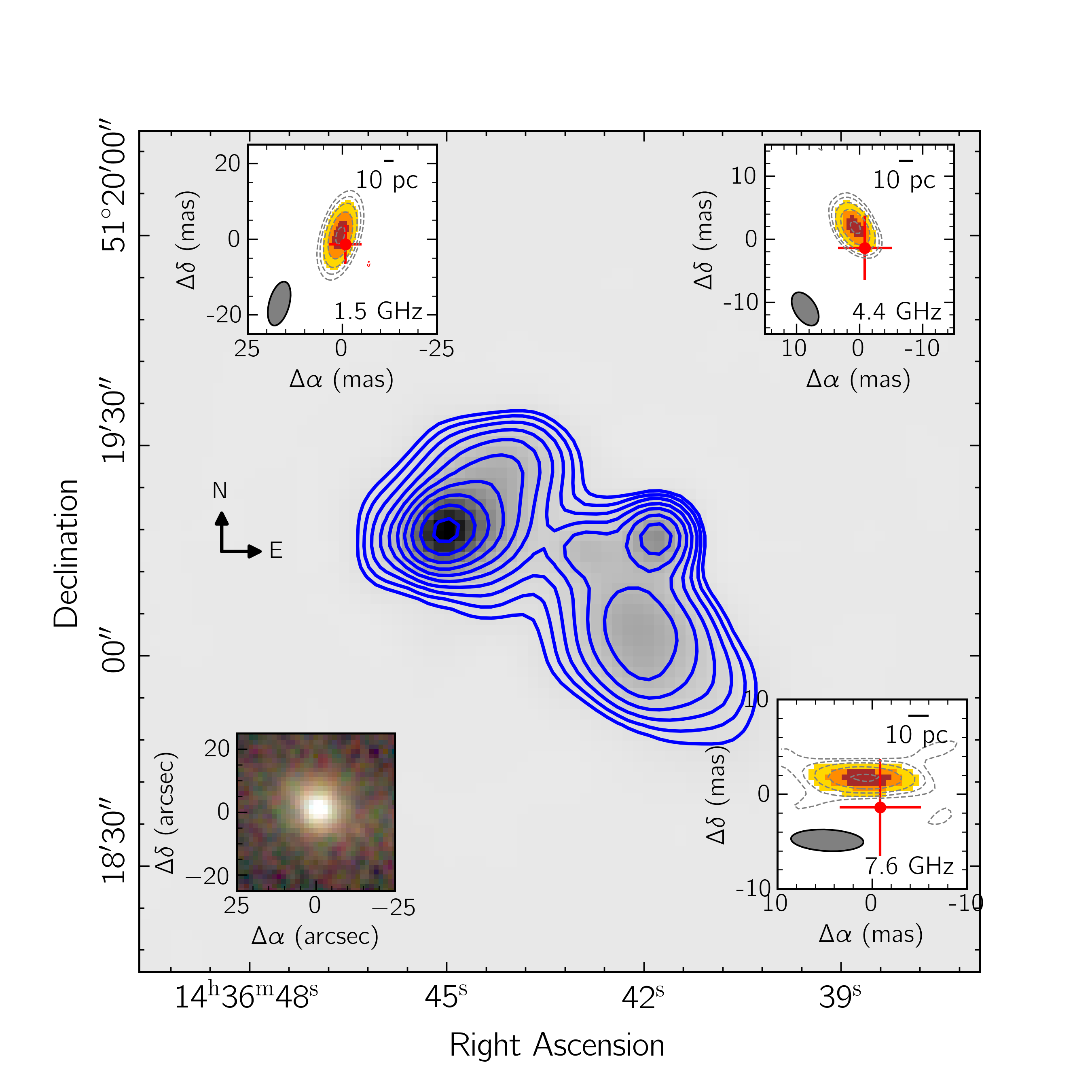}\\
\includegraphics[width=.49\linewidth]{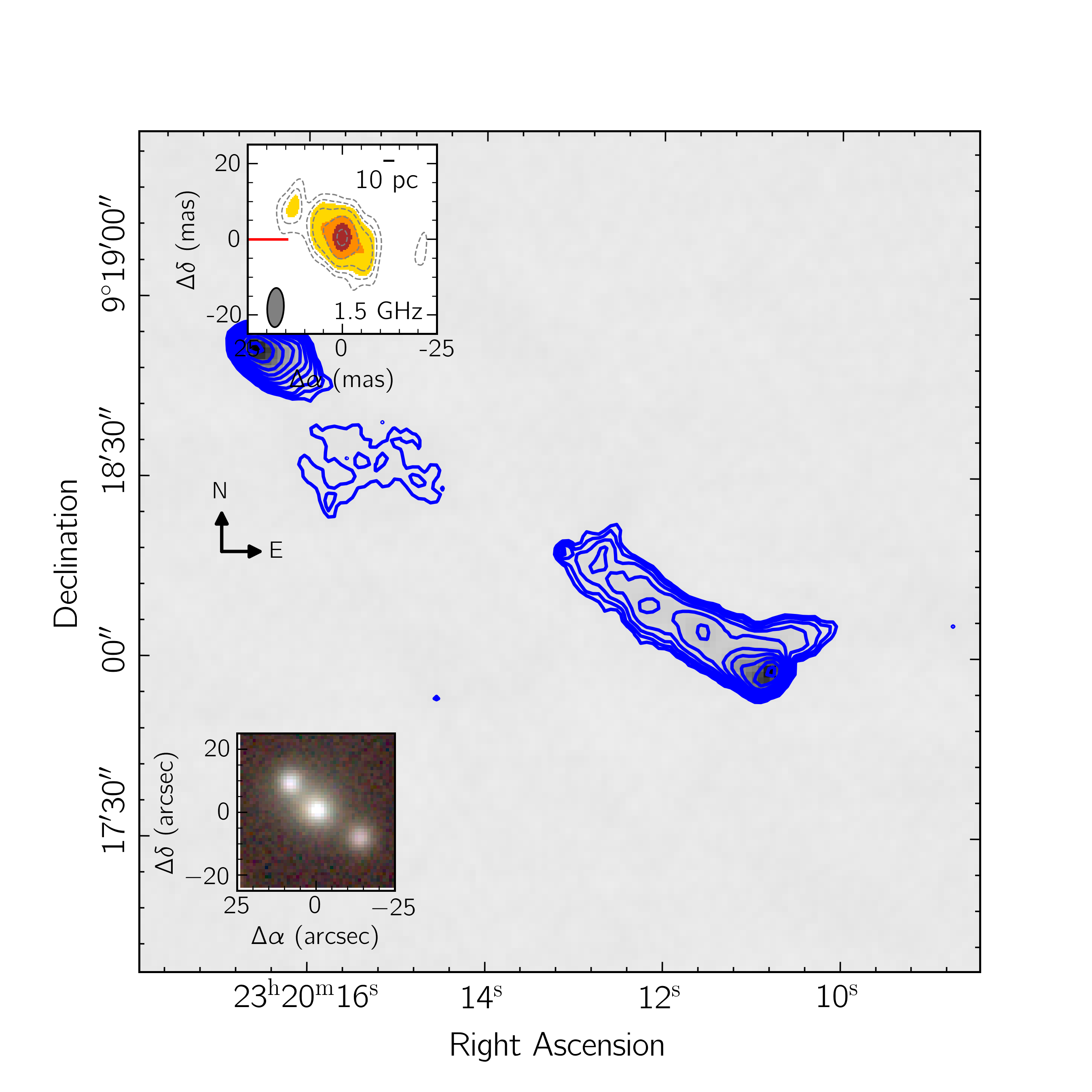}
\includegraphics[width=.49\linewidth]{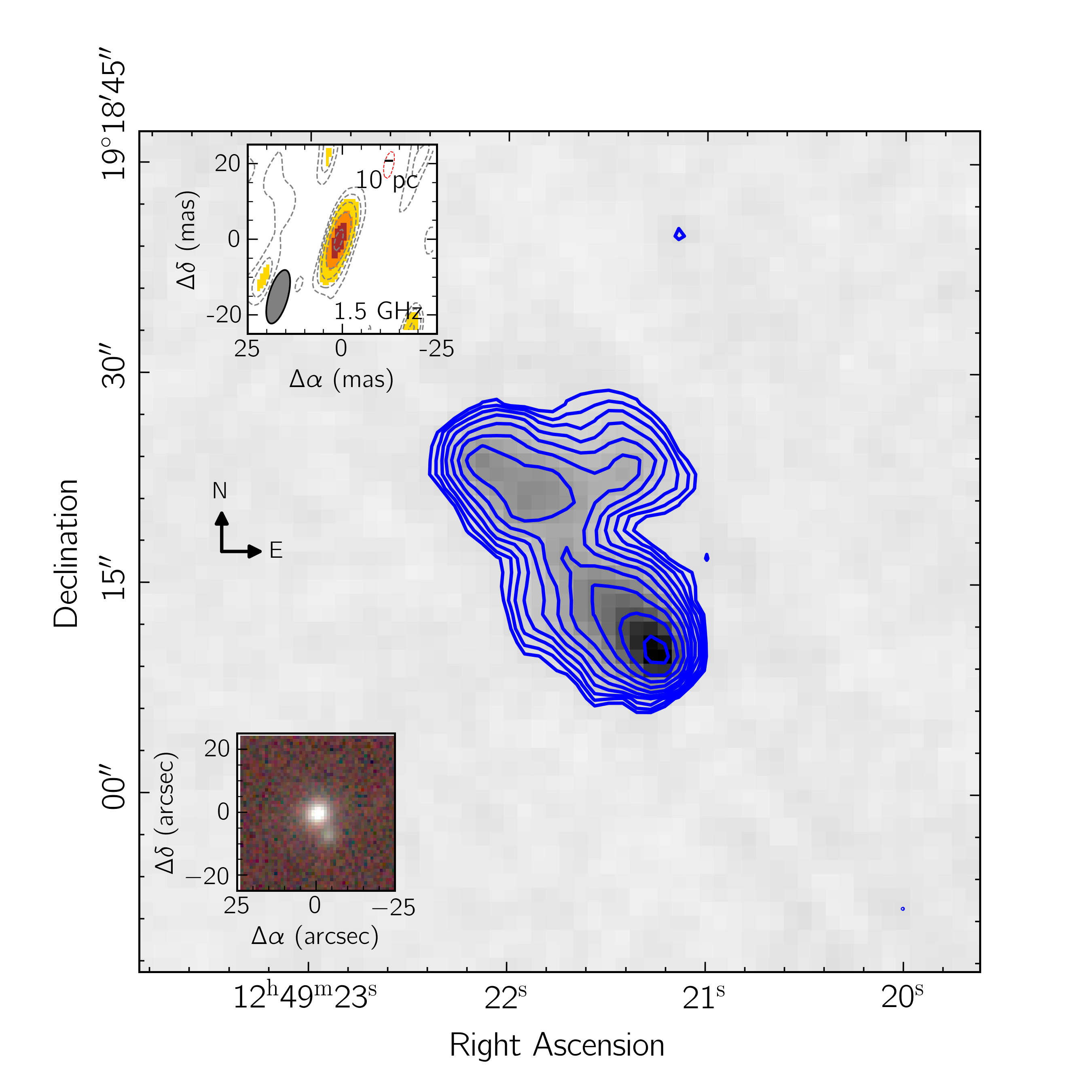}
\caption{The extended kpc scale radio morphology for 3C 223.1 using 144 MHz LOFAR ({\it top left}), and maps from the 2.5 GHz  VLASS survey \citep{2020PASP..132c5001L} for J1430+5217 ({\it top right}), J2315+1027 ({\it bottom left }) and TXS 1244+200 ({\it bottom right}) respectively.  The insets show the 
parsec-scale VLBA radio core in available 1.5 GHz, 4.3 GHz, or 7.6 GHz maps and respective beam FWHM with contour levels set on $\sigma \, \times [0.1, 0.2, 0.3, 0.5, 1, 2, 4, 7, ...]$ in log scale. The Gaia DR3 positions are indicated by red crosses, with their lengths representing the 1$\sigma$ rms errors. The bottom-left inset shows a DECaLS ($g,r,z$) color composite image centered on the radio host galaxy.}
\label{fig:VLBA_XRGs}
\end{figure*}

\subsubsection{J2315+1027}
Among the 7 XRGs with VLBA counterparts, we detect a resolved central core for only J2315+1027, in our 1.4 GHz VLBA maps (see Figure~\ref{fig:VLBA_XRGs}). At $z = 0.255$, the beam size of 10.4 milliarcsec (mas) offers a spatial resolution of 40 pc. Interestingly, the resolved cores are seen at a separation of $\sim $40 pc. (see Fig.~\ref{fig:VLBA_XRGs}). At parsec scales, the core is found to extend at a position angle (PA) of $\rm 61^\circ$, measured from north to east direction. It aligns with the primary kiloparsec-scale jet observed at a PA of $\rm 61.2^\circ$ in the VLA FIRST observations. A follow-up high-frequency VLBA observation constraining the spectral index would help to verify the possibility of a binary black hole or core-jet system. Interestingly, the kpc-scale active FRII-jet is aligned towards the major axis of the optical host galaxy. This is in agreement with the observed tendency of active jets in XRGs to align along the major axis of their optical host galaxies  \citep{2019ApJ...887..266J, Giri2026arXiv260405471G}. Further, the DECaLS optical field reveals three close companion galaxies with photometric redshifts near that of the XRG, suggesting a potential merger scenario (see Figure~\ref{fig:VLBA_XRGs}). 

\subsubsection{3C 223.1}
3C 223.1, is known to have a kpc-scale ``double-boomerang" morphology. Recently, \citet{2022A&A...663L...8G} have shown that the wings anomalously exhibit a flatter radio spectrum even compared to the hotspots in the primary lobes, which supports the particle acceleration associated with the rebounding of collimated backflow of synchrotron plasma streaming through the primary lobes. Our 1.4 GHz VLBA maps provide a hint of a resolved structure at a PA of $22^\circ$, which is roughly aligned toward the active radio jets having a PA of $11^{\circ}$. The resolved jet structure is further supported by the 5 GHz VLBA maps by \citet[][ see their figure 4]{2024MNRAS.530.4902S} showing an extended pc-scale (PA of $25^\circ$) morphology, likely indicating a core jet structure. Furthermore, a relatively flat spectral index of $\alpha = 0.32$, derived from the 1.4 GHz and 5 GHz VLBA maps, in conjunction with the presence of double-peaked emission lines in the AGN locus of the BPT diagram, suggests that this source is a promising candidate for follow-up observations to search for a binary AGN.

For the remaining five sources, the core is unresolved, with a typical spatial beam size resolution ranging between 8.7 to 82 pc. Based on multi-frequency VLBA observations for these 5 sources, we estimate spectral indices ranging from 0.01 to 0.54 (see Table~\ref{table:vlbaxrg}, column 9). This suggests that, if XRGs with DPNELs host binary SMBHs, their separation must be smaller than the beam size.

\begin{figure*}[!htbp]
\centering
\includegraphics[width=.33\linewidth]{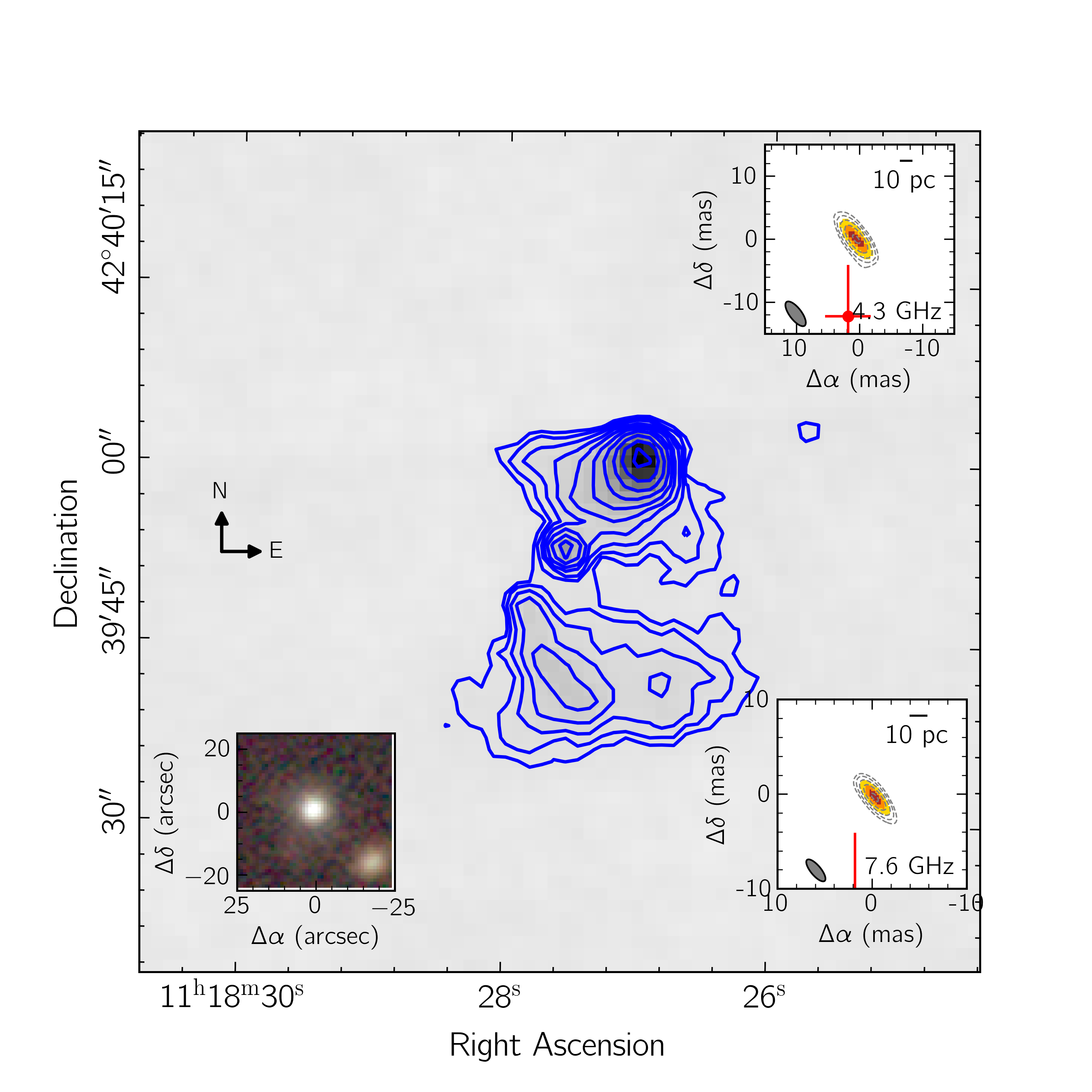}
\includegraphics[width=.33\linewidth]{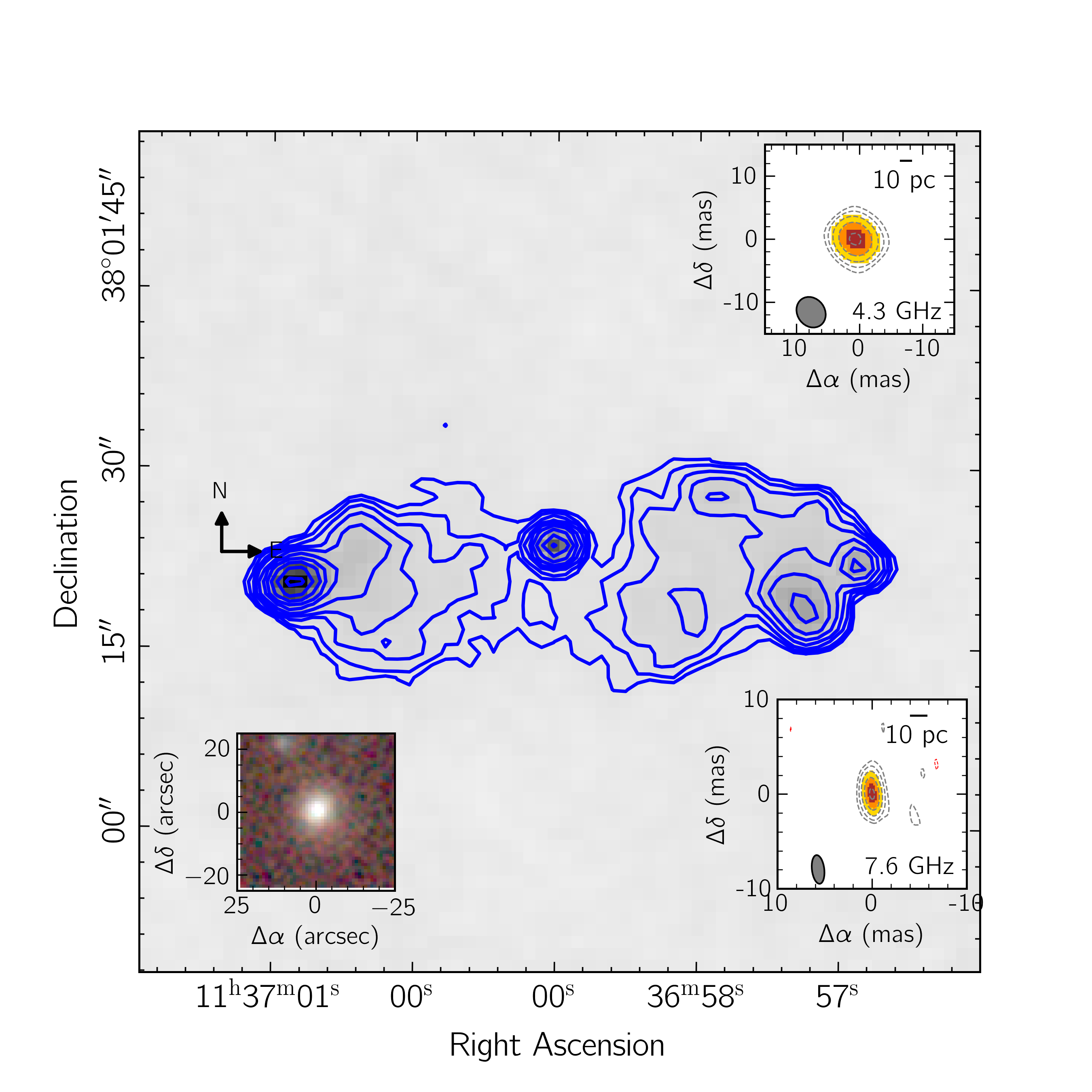}
\includegraphics[width=.33\linewidth]{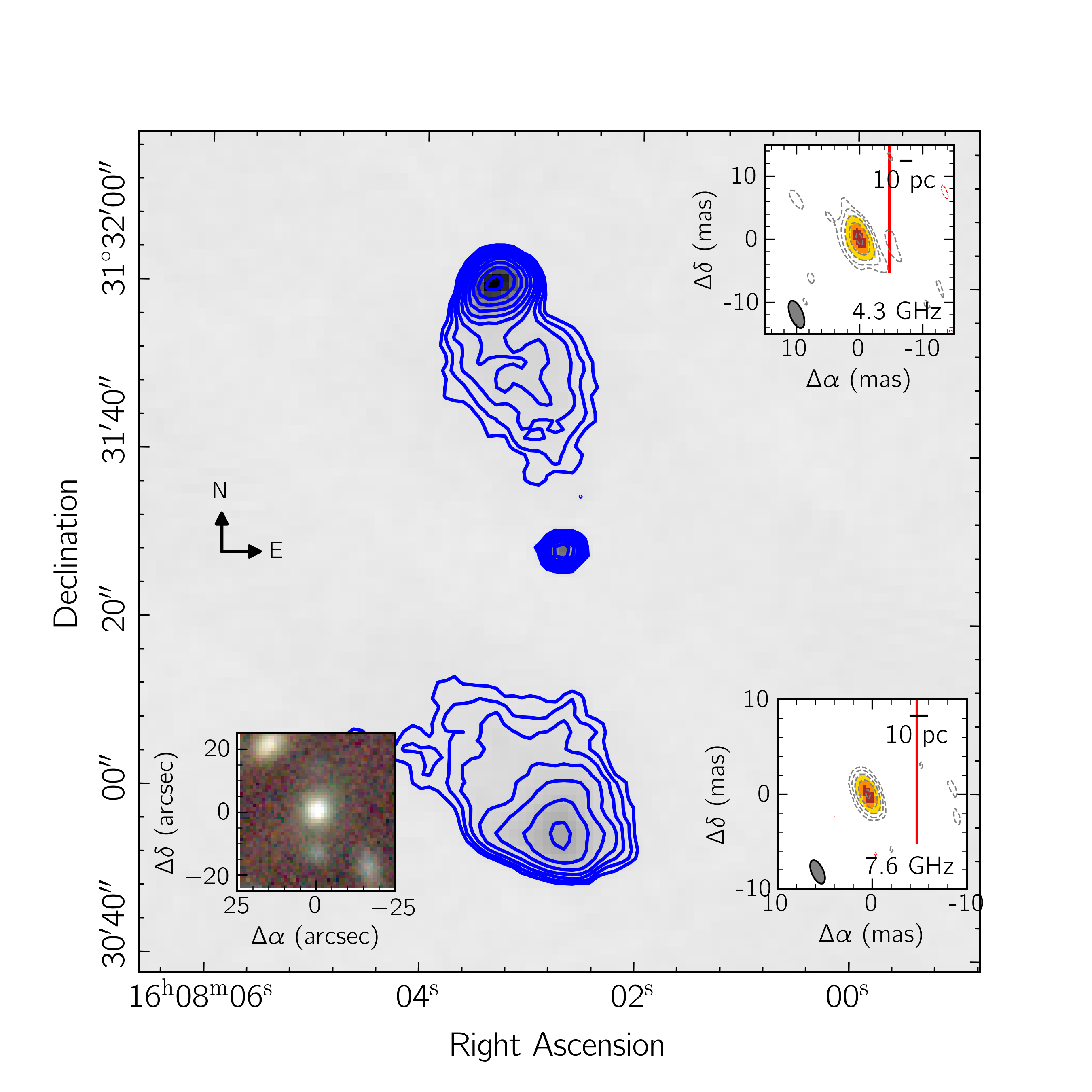}
\caption{As in Figure~\ref{fig:VLBA_XRGs} with 2.5 GHz VLASS maps showing extended kpc scale radio morphology for J1115+4314, J1134+3835 and J1604+3221 (left to right).}
 \label{VLBI_XRGs_EXTRA_1}
\end{figure*}

\subsubsection{Gaia-VLBA offsets}
We may recall that the positional offset of radio-optical emission cores may signify the presence of multiple AGNs \citep{2023ApJ...958...29C,2024ApJS..274...28C, 2024ApJ...969...36X}. To test the multi-AGN hypothesis, which we have seen is hinted by multiple indirect indicators, notably DPNEL, X-shaped morphology, and flat-spectrum core, we now compare the spatial positions of VLBA radio maps with those from Gaia DR3 \citep{2023A&A...674A...1G}. We obtain a significant offset (between 1.5 to 33 mas) for 5 out of 7 sources. A flat-spectrum core along with a radio-optical offset in XRGs J1430+5217, J1115+4314, and J1604+3221, makes them strong binary black hole candidates \citep{2013A&A...553A..13O}. Even though significant offsets are detected in the multifrequency observations of 3C 223.1 and J2315+1027, their spectral indices are used to confirm the possibility of dual AGN. For the above sources, the Gaia optical counterparts are offset towards the resolved (or hint of resolved) VLBI components. Such offsets can plausibly arise from a radio-loud core with optical emission rendered undetected (due to obscuration) and a radio-quiet secondary core dominating the emission detected by Gaia. The majority of recent Gaia–VLBI studies at mas or sub-mas level show an offset that could arise from strong optical jet emission preferentially being located either downstream or upstream of the peak of the AGNs jet radio emission \citep{2017A&A...598L...1K,2019MNRAS.485.1822P}. Therefore, other explanations, such as a single core driving both optical and radio emission with a significant separation, or a frequency-dependent core position shift due to flaring, cannot be discounted \citep{2011A&A...532A..38S,2019MNRAS.485.1822P}. 

\subsubsection{Role of NLR in occurrence of DPNELs}
We further explore the binary black hole scenario, where the DPNELs are considered to be originating from the separate narrow line regions (NLR) of the individual SMBHs. Following \citet{2013ApJ...762..110L}, we determine the size of the NLR as: 
\begin{equation}
\label{eqn:radius_nlr}
\begin{split}
\log\left(\frac{\rm R_{NLR}}{\rm pc}\right) = \, & (0.250 \, \pm \, 0.018) \times  \log \left( \frac{\rm L_{[OIII]}}{10^{42} \rm \ ergs \, s^{-1}}\right) \, \\
& + (3.746 \,  \pm \, 0.028)
\end{split}
\end{equation}
where $\rm L_{[OIII]}$ represents the \oiiib\ luminosity, corrected for the Balmer extinction of the individual emission line components. In the above scenario, one expects the separation of individual NLRs to be smaller than the BH separation. The size of NLR is listed in columns (6) and (7) of Table~\ref{table:xrg}. The NLR size for DPNEL XRGs ranges between 1.52 and 17.74 kpc, which is significantly larger than the expected binary black hole separation of 0.1 kpc \citet[see also][]{2015AJ....149...92V}. Nonetheless, considering the extended structure of the NLR along with the multiple double-peak emission lines (\hbeta, and \halpha) which trace the same kinematics as \oiiiab, \niiab and \siiab, the observed DPNEL in the XRGs could still be associated with the binary black hole system \citep{2009ApJ...705L..20X,2009ApJ...704.1189T}.  It is worth reiterating that the DPNELs can also arise from alternative processes, notably complex kinematics within the NLR, such as rotating disks and biconical outflows  (see Section~\ref{sec:intro}). Follow-up narrow-band imaging and IFS observations would help to distinguish the various possibilities.

\begin{figure*}[!htbp]
\centering
\includegraphics[width=0.99\textwidth]{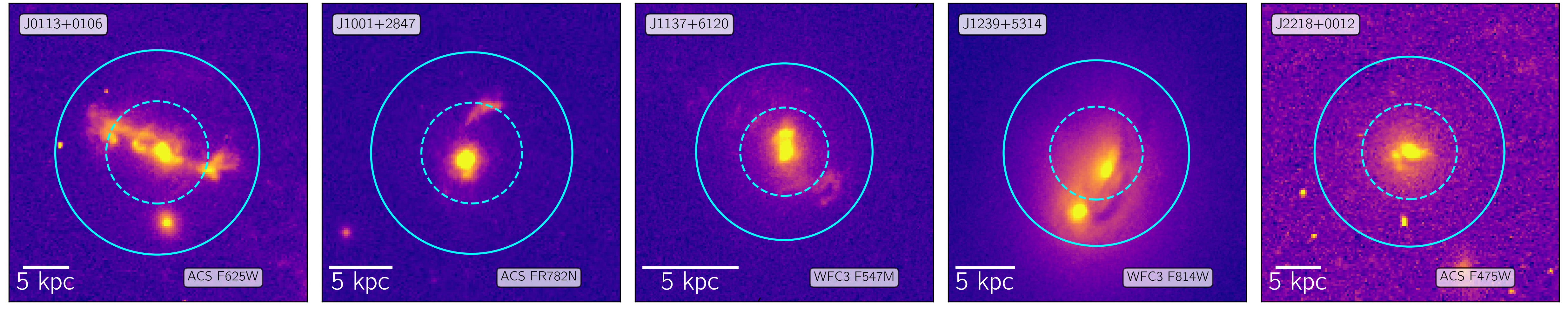}
\caption{The high-resolution HST/WFC3  images of XRGs with DPNELs revealing signatures of galaxy mergers and distinct tidal tails, indicative of an ongoing or recent galactic merger. The SDSS fiber of 3 arcsec {\it (blue solid circle)} and the DESI fiber of 1.5 arcsec diameter {\it (blue dashed circle)}, encompassing the merging galaxies, capture the double peak emission lines, suggesting the presence of a likely dual AGN system.
}
\label{fig:merger}
\end{figure*}

\subsection{Double-peaked XRGs and the role of galaxy merger}

High-resolution hydrodynamical simulations of the NLR regions in the AGNs hosted by merging galaxies suggest the presence of DPNELs in the late stages of the merger \citep{2013MNRAS.429.2594B}. Probably only a small fraction of double peaks are directly relatable to the relative motion of the SMBHs, while most originate from complex gas kinematics near the SMBH. Nevertheless, around 80\% of these profiles show some influence of SMBH dynamics, manifested as shifted peaks or velocity offsets. Among the merger-induced DPNELs, approximately 10–40\% result in SMBH separations in the range of 0.1 to 1 kpc. Furthermore, \cite{2017MNRAS.469.4437C} explored AGN triggering in merging systems by varying parameters such as galaxy mass ratios, orbital configurations, gas content, and black hole properties, concluding that dual AGNs are found in up to 20–30\% of major mergers and 1–10\% of minor mergers. This is further supported by \cite{2023MNRAS.524.4482B}, where, using automated techniques, 159 dual AGNs are detected within $z \le 0.75$ and are mainly associated with red and evolved galaxies, indicating a merger origin. In contrast, the follow-up optical long-slit spectroscopy by \cite{2018ApJ...867...66C} of 95 SDSS galaxies exhibiting double-peaked narrow AGN emission lines revealed that only eight sources ($<$10\%) possess companion galaxies with line-of-sight velocity separations below 500 km/s and projected physical separations under 30 kpc.

Given that $\sim 95\%$ of double-peaked emission line XRGs likely host dual AGNs, and considering the galaxy overdensity with similar color in proximity to the central radio source in deep DECaLS images, we further explore the role of galaxy mergers in the formation of XRGs. To identify potential mergers, we search for galaxies located near the central radio core that have potential companion matching spectroscopic redshifts within $\Delta z = 0.003$, corresponding to a peculiar velocity of $\sim 1000 \rm \ km\ s^{-1}$  in small groups \citep{1978ApJ...222..784G, 1993ApJ...404...38G}. Here, we examine galaxies within a 30-arcsecond radius, which corresponds to $\sim$150 kpc at the average redshift of 0.3, well within the typical virial radius of a central galaxy hosting an AGN. In our study, we crossmatch these galaxies with DECaLS DR10 to obtain photometric redshifts derived using a random forest algorithm \citep{2023JCAP...11..097Z}, while spectroscopic redshifts are taken from the SDSS and DESI samples. For our sample of \oiiidp DPNEL XRGs, 17 ($\sim$30\%) exhibit at least one companion galaxy likely indicative of a merger event, whereas 2 XRGs are associated with multiple companion galaxies. The fraction of multiple galaxies rises to $\sim$45\% (26/58) if we consider the photo-$z$ selected systems within $\Delta z = 0.05$.  The galaxy overdensity around XRGs is further supported by \citet{2019ApJ...887..266J}, where a higher median clustering richness  (quantified by the number of SDSS galaxies with $M_r \geq -19$ located within a projected radius of 1 Mpc and a redshift interval of $\Delta z = 0.04(1+z)$) of 8.9 is found around 107 XRGs at $z < 0.4$. The excess merger fraction further suggests that the large NLR sizes estimated from the central black hole (Eqn.~\ref{eqn:radius_nlr}) may possibly be due to the fact that empirical relations may not hold true for merger systems. Interestingly, among the total sample of the \totxrgs XRGs,  all five with DPNEL that have deep HST high-resolution images available reveal a close companion and show strong merger signatures with tidal tails (see Figure~\ref{fig:merger}). The SDSS as well as DESI fiber spectra integrate light from both galaxies, resulting in a double-peaked emission profile. This highlights the possible contribution of mergers, which may produce jet reorientation, leading to X-shaped morphology, and making them probable hosts for dual/binary SMBHs as well \citep{Giri2026arXiv260405471G}.

\section{Conclusions}
In this study, we performed a systematic search for dual AGNs among \totxrgs XRGs with available SDSS and DESI spectra. Additionally, we report the parsec-scale morphology of several double-peaked XRGs using radio VLBA maps, leading to the following key findings:

\begin{enumerate}

\item The XRGs preferentially show a higher (30\%) occurrence rate of double-peaked narrow emission lines than the rate of just $\sim1\%$ found for the general population of galaxies. In addition, the probability of both emission line components in the DPNEL profile exhibiting AGN nature, based on the BPT diagnostic diagram of both [N{~\sc ii}] and [S{~\sc ii}] lines, is three times higher in XRGs ($\sim 95\%$) than in the control sample of non-XRGs (22\%). This is further corroborated by various MIR color excess and diagnostics in XRGs, yielding an AGN fraction of 91\%. This indicates a higher probability of XRGs hosting dual/binary AGNs. However, given the strong radio jets in XRGs, the excess DPNELs fraction may also arise from alternative mechanisms, including jet–ISM interactions and outflows.

\item The fraction of DPNEL galaxies hosting dual-AGN signature is found to depend strongly on the radio luminosity. This fraction increases from 25\% for the radio-undetected to 54\% in radio-detected galaxies. The $\sim$95\% likelihood of dual AGN candidates, seen as nebular emission line ratios induced by AGN-like photoionisation in radio-bright DPNEL XRGs and FR-II radio galaxies, supports the merger-driven scenario for the formation of dual/binary AGN.

\item  The radio core is resolved for one XRG, J2315+1028, in the parsec-scale VLBA maps. The resolved/extended structure reported here is collinear with the jet direction, indicating either a core-jet configuration or a dual AGN scenario. For 5 of 6 XRGs with unresolved VLBA core, the spectral index between 4 GHz and 8 GHz, ranging from 0.01 to 0.54, indicates a flat spectrum for the central AGN. The mas-scale radio-optical offset between VLBA and Gaia detections also signifies the presence of multiple cores. In conjunction with both the double-peak emission line components exhibiting an AGN nature, radio cores with flat spectral indices, and radio-optical offsets between the VLBA and Gaia positions, these features are strongly indicative of XRGs as dual/binary AGN candidates.

\item The double-peaked XRGs harbor similarly massive black holes and have similar radio luminosities as single-peaked XRGs. We find that more than 30\% of DPNEL XRGs have a spectroscopically matched companion galaxy, suggesting that mergers play a substantial role in the formation of X-shaped radio morphology. 

\section*{Acknowledgements}
The authors sincerely thank the anonymous reviewer for their valuable comments and suggestions, which have helped us improve the quality of this manuscript. GP acknowledges the Indian Institute of Astrophysics (IIA) for the opportunity to carry out this work under the Visiting Student Program. GP and KM acknowledge the support of the Polish National Science Center (NCN) grant UMO-2024/53/B/ST9/00230. LCH was supported by the National Science Foundation of China (12233001) and the China Manned Space Program (CMS-CSST-2025-A09). SM acknowledges Ramanujan Fellowship (RJF/2020/000113) by DST-ANRF, Govt. of India for this research. VN is supported by Beijing Natural Science Foundation (Grant No. IS25004). GK thanks Indian National Science Academy for a Senior Scientist position during which this project started. \par

Funding for the Sloan Digital Sky Survey V has been provided by the Alfred P. Sloan Foundation, the Heising-Simons Foundation, the National Science Foundation, and the Participating Institutions. SDSS acknowledges support and resources from the Center for High-Performance Computing at the University of Utah. SDSS telescopes are located at Apache Point Observatory, funded by the Astrophysical Research Consortium and operated by New Mexico State University, and at Las Campanas Observatory, operated by the Carnegie Institution for Science. The SDSS website is \url{www.sdss.org}.SDSS is managed by the Astrophysical Research Consortium for the Participating Institutions of the SDSS Collaboration, including Caltech, The Carnegie Institution for Science, Chilean National Time Allocation Committee (CNTAC) ratified researchers, The Flatiron Institute, the Gotham Participation Group, Harvard University, Heidelberg University, The Johns Hopkins University, L'Ecole polytechnique f\'{e}d\'{e}rale de Lausanne (EPFL), Leibniz-Institut f\"{u}r Astrophysik Potsdam (AIP), Max-Planck-Institut f\"{u}r Astronomie (MPIA Heidelberg), Max-Planck-Institut f\"{u}r Extraterrestrische Physik (MPE), Nanjing University, National Astronomical Observatories of China (NAOC), New Mexico State University, The Ohio State University, Pennsylvania State University, Smithsonian Astrophysical Observatory, Space Telescope Science Institute (STScI), the Stellar Astrophysics Participation Group, Universidad Nacional Aut\'{o}noma de M\'{e}xico, University of Arizona, University of Colorado Boulder, University of Illinois at Urbana-Champaign, University of Toronto, University of Utah, University of Virginia, Yale University, and Yunnan University. The VLBA is operated by the National Radio Astronomy Observatory, a facility of the National Science Foundation operated under cooperative agreement by Associated Universities, Inc. The VLBA observation of 0932+075 at 15.4 GHz carried out on 22 March 2001 was proposed by the CLASS team, specifically by Neal J. Jackson and Martin Norbury. It was devised to be an ultimate proof that 0932+075 was not a GL system. However, the image resulting from that observation has never been published. The raw data belongs to the public domain yet we contacted the proposers and we have received written permission to publish the map resulting from that data in the present paper. This research has made use of the NASA/IPAC Extragalactic Database (NED) which is operated by the Jet Propulsion Laboratory, California Institute of Technology, under contract with the National Aeronautics and Space Administration.
\end{enumerate}


\bibliography{apj}{}
\bibliographystyle{aasjournalv7}



\end{document}